\documentclass[pra,showpacs,preprint]{revtex4}

\usepackage{epsfig}
\usepackage{amsmath}
\usepackage{bbm}

\begin{document}
\title{Controlled exchange interaction
for quantum logic operations with spin qubits
in coupled quantum dots}

\author{S. Moskal}
\affiliation{Faculty of Physics and Applied Computer Science, AGH
University of Science and Technology, Krak\'ow, Poland}
\author{S. Bednarek}
\thanks{Electronic address: bednarek@novell.ftj.agh.edu.pl}
\affiliation{Faculty of Physics and Applied Computer Science, AGH
University of Science and Technology, Krak\'ow, Poland}
\author{J. Adamowski}
\affiliation{Faculty of Physics and Applied Computer Science, AGH
University of Science and Technology, Krak\'ow, Poland}

\date{\today}

\begin{abstract}
A two-electron system confined in two coupled semiconductor quantum dots
is investigated as a candidate for performing quantum logic
operations on spin qubits.
We study different processes of swapping the electron
spins by controlled switching on/off the exchange interaction.
The resulting spin swap corresponds
to an elementary operation in quantum information processing.
We perform a direct time evolution simulations
of the time-dependent Schr\H{o}dinger equation.
Our results show that -- in order to obtain the full interchange
of spins -- the exchange interaction
should change smoothly in time.
The presence of jumps and spikes in the corresponding
time characteristics leads to a considerable increase
of the spin swap time.  We propose several
mechanisms to modify the exchange
interaction by changing the confinement potential profile
and discuss their advantages and disadvantages.
\end{abstract}

\pacs{03.67.Lx,03.67.Mn,73.21.La}

\maketitle

\section{Introduction}
In quantum computation, one and two-qubit logic gates play
a crucial role, since they are allow us to perform
arbitrary quantum logic operation \cite{DiVincenzo1995}.
Recently, a practical realization of these gates is a challenge
for many physical laboratories.  Among several propositions
of constructing the systems performing the two-qubit gates,
the most promising are these based on the mechanism
of controlled switching on/off the exchange interaction
between spin qubits
\cite{Kane1998,Loss1998,Imamoglu1999,Vrijen2000,Engel2001,Levy2001,%
Leuenberger2001,Privman2002,Friesen2003,Stoneham2003,Feng2003,Engel2004,Elzerman2005}.
Tuning the exchange interaction between electrons
in coupled quantum subsystems can lead to the interchange
of qubits and the performance of a designed quantum
logic operation.

A physical implementation of the controlled exchange interaction
is itself a difficult task.  Moreover, the elements of
a future quantum computer
have to fulfill several conditions, known in the form
of DiVincenzo criteria \cite{DiVincenzo1998}.
One of them is a scalability, which allows to join
the elements into a large computational machine.
It is expected that semiconductor nanodevices, in particular
those consisting semiconductor quantum dots (QDs),
should be scalable to large enough size.
Recently, the electron spin states
are the most promising candidates for qubits
\cite{Engel2001,Leuenberger2001,Privman2002}
due to the long coherence time
\cite{Hanson2003}.
The quantum information processing can be performed
via the changes of electron spin states
\cite{Leuenberger2001,Burkard1999}.
The spin of the electron can be rotated as a result of
precession in the external static magnetic field
\cite{Sasakura2001,Elzerman2005}.
The spin swapping can also be obtained
when irradiating the electron system by
the electromagnetic wave with the frequency
adjusted to the Zeeman splitting in the
external magnetic field
\cite{Vrijen2000,Wu2004}.

In the present paper, we study the process of swapping
the electron spins,
which results from the controlled switching of the exchange interaction
between the electrons in coupled QDs.
The tuned exchange interaction should lead to the rotation
of electron spin
\cite{Kane1998,Loss1998,Privman2002,Elzerman2005,Burkard1999,DiVincenzo2002}.
The theoretical model proposed takes into account
all the three space dimensions and the electron spins.
We apply the adiabatic approximation in order to
decouple the transverse and longitudinal
degrees of freedom and perform the integration
over the transverse spatial coordinates.  This leads
to the effectively one-dimensional two-electron problem
\cite{Bednarek2003eff}, which can be solved by the numerical
method with an arbitrary precision \cite{ITS}.
The system under study is described by the two-particle
wave function of the form of four-component state vector,
which takes into account all possible spin configurations.
We consider different methods of turning on and off the exchange
interaction between the electrons and investigate
the resulting rotation of electron spins.
The present computer simulations are based
the accurate numerical solutions of the time-dependent
Schr\H{o}dinger equation, i.e., they allow us to trace
the time evolution of the electron system in a direct manner.
Therefore, the present results fill in a gap between
the quantum information theory and experiment.
We hope that these results will serve as a guide
for the experimental groups who are involved
in designing and constructing the nanodevices
with spin qubits.

The paper is organized as follows:
the theoretical model is presented in Sec. II,
the numerical method is described in Sec. III,
the results are presented in Sec. IV (for the vertically
coupled QDs) and in Sec. V (for the laterally coupled QDs).
Section VI contains conclusions and summary.

\section{Theoretical model}

The model applied in the present paper is an extension of that
proposed by the present authors in Ref. \cite{Moskal2005}.
Contrary to Ref. \cite{Moskal2005},
we explicitly take into account the spin states of electrons.
For sake of completeness, we repeat below the major
steps of the approach \cite{Moskal2005}.
We study the two electrons localized in the double QD
nanostructure.
We assume the cylindrical symmetry of
the system with symmetry axis ($z$ axis) going
through the geometrical centers of the QDs.
The potential confining the electrons
in the $x-y$ plane is taken to be so sufficiently strong
that the differences between the energy levels resulting from
the $x-y$ space quantization are much larger
than the energy of spin swapping.
This potential is usually
called the lateral confinement potential.
We approximate the lateral potential by the two-dimensional
harmonic oscillator potential and assume
that both the electrons occupy
the ground state in the $x-y$ motion.
In other words,  we assume that the lateral electron degrees of freedom
are frozen.  The above assumptions allow us to reduce
the starting three-dimensional problem
to the effectively one-dimensional problem
\cite{Bednarek2003eff} with the electron-electron
interaction being the following function
of $z_1$ and $z_2$ coordinates of both the electrons:
\begin{equation}
\label{Ueff}
U_{e\!f\!f}(z_{1}-z_{2})=\frac{e^{2}\sqrt{\pi\beta}}
{4\pi\varepsilon_{0}\varepsilon}
e^{\beta|z_{1}-z_{2}|^{2}}
\mathrm{erfc} \left(\sqrt{\beta}|z_{1}-z_{2}|\right) \; ,
\end{equation}
where $\varepsilon_0$ is the vacuum electric permittivity,
$\varepsilon$ is the static electric permittivity,
$\beta = m_e \hbar\omega_{\perp} /2$, $me$ is the electron
effective conduction band mass,
and $\hbar\omega_{\perp}$  is the excitation energy levels of the electron
transverse motion.
The Hamiltonian of the system takes the form
\cite{Bednarek2003eff}
\begin{equation}
\label{Hamil}
H = -\frac{\hbar^{2}}{2m_e}
\left( \frac{\partial^2}{\partial z_1^2}+\frac{\partial^2}{\partial z_2^2}\right)
+ U(z_1)+U(z_2)+U_{e\!f\!f}(z_{1}-z_{2})+2\hbar\omega_{\perp} \; ,
\end{equation}
where $U(z_{i})$ is the potential energy of the vertical confinement
of the $i$th electron.  In the following,
vertical confinement $U(z)$ will be taken
in a form of two potential wells separated by the
potential barrier.
The energy is measured with respect to the conduction band bottom
of the quantum well material.
In the calculations, we take on the parameters, which correspond to
nanostructure based on GaAs: $m_e=0.067 m_{e0}$,
where $m_{e0}$ is the free electron rest mass, and
the static electric permittivity $\varepsilon=11.0$.

In the two-electron system, we are dealing with four independent
spin states; therefore, the total wave function of the system
can be represented by the following four-component vector:
\begin{equation}
\label{Psivec}
\Psi(z_1,\sigma_1,z_2,\sigma_2) = \left(
\begin{array}{c}
\psi_{\uparrow\uparrow}   \\
\psi_{\uparrow\downarrow} \\
\psi_{\downarrow\uparrow} \\
\psi_{\downarrow\downarrow}
\end{array} \right) \; ,
\end{equation}
where $\sigma_j$ ($j=1,2$) are the spin coordinates
of the electrons, $\psi_{nm} = \psi_{nm}(z_1,z_2)$
are the basis wave functions with the indices
$n$ and $m$ = $\uparrow$ and $\downarrow$, which
correspond to the $z$-spin component eigenvalues
$+\hbar/2$ and $-\hbar/2$.
Basis wave functions $\psi_{nm}$
do not possess a well defined symmetry.
However, total wave function (\ref{Psivec})
has to be antisymmetric with respect
to simultaneous exchange of the space and spin coordinates.
This property imposes the following conditions
on the basis wave functions:
\begin{subequations}
\label{psi_cond}
\begin{eqnarray}
\psi_{\uparrow\uparrow}(z_1,z_2) &
= & - \, \psi_{\uparrow\uparrow}(z_2,z_1) \;, \\
\psi_{\downarrow\uparrow}(z_1,z_2) &
= &  - \, \psi_{\uparrow\downarrow}(z_2,z_1) \;, \\
\psi_{\downarrow\downarrow}(z_1,z_2) &
= & - \, \psi_{\downarrow\downarrow}(z_2,z_1) \; .
\end{eqnarray}
\end{subequations}
In representation (\ref{Psivec}), the operators of the total spin
are expressed by the following $4\times 4$ matrices:
\begin{equation}
\label{sigma_tot}
\sigma^{tot}_x =
\left(\begin{array}{cc}
\sigma_x & \mathbbm{1}  \\
\mathbbm{1} & \sigma_x  \\
\end{array} \right) \; ,
\quad \sigma^{tot}_y =
\left( \begin{array}{cc}
\sigma_y & -i\mathbbm{1}  \\
i\mathbbm{1} & \sigma_y  \\
\end{array} \right) \;,
\quad \sigma^{tot}_z =
\left( \begin{array}{cc}
\sigma_z + \mathbbm{1} & 0  \\
0 & \sigma_z - \mathbbm{1}  \\
\end{array} \right) \; ,
\end{equation}
where $\mathbbm{1}$ denotes the unit 2$\times$2 matrix
and $\sigma_x, \sigma_y, \sigma_z$ are the spin Pauli matrices.
The corresponding expectation values are calculated
as $\langle\sigma^{tot}_k\rangle =
\langle\Psi\sigma^{tot}_k \Psi\rangle$,
where $k=x,y,z$.  Multiplying $\langle\sigma^c_k\rangle$
by $\hbar/2$ we obtain the expectation values
of the components of the total spin, i.e.,
$S^{tot}_k = (\hbar/2)\langle\sigma^{tot}_k\rangle $.

Due to their indistinguishability the electrons
can not be numerated; therefore, we can not determine
which electron is in a given spin state.
However, we can distinguish the two quantum dots,
determined by the corresponding potential wells.
We shall call them the left (L) and right (R)
quantum dot (potential well).
If both the QDs are separated by the potential barrier,
which prohibits a tunneling of electrons, then in the ground state
of the system, each electron is localized in a single QD.
Owing to this, we can determine the spins of electrons
in the left and right QD.  For this purpose, we introduce
the auxiliary wave function
$\varphi(z_1,z_2)$, which will be called the reference wave function.
This wave function is a solution of the Schr\H{o}inger equation,
for the two distinguishable particles, which
-- with an exception of indistinguishability -- possess
all the properties of the electrons.
For the distinguishable particles we do no perform a symmetrization
of the two-particle wave function.  Instead the solutions
found for $\varphi(z_1,z_2)$ correspond to the configurations,
in which one electron (described by coordinate $z_1$) is localized
in the left QD and the other ($z_2$) is in the right QD.
This state always exists if the tunneling through the barrier
is not possible.  In each simulation performed,
these configurations are realized at the initial ($t=0$)
and final ($t=T$) time instants.
The reference wave function will serve to a construction
of the initial state wave function and a determination
of the spin states of electrons in both the QDs.
The expectation values of spin components are calculated
as follows
\begin{equation}
\label{spin1}
S^j_k = \frac{\hbar}{2}\langle{\sigma^{j}_k}\rangle
= \frac{\hbar}{2} u^T \sigma^{j}_k u \;,
\end{equation}
where  $j=L,R$ corresponds to the electron localized
in the left (L) and right (R) QD, and
\begin{equation}
\label{u}
u = \left(
\begin{matrix}
\langle\varphi|\psi_{\uparrow\uparrow}\rangle   \\
\langle\varphi|\psi_{\uparrow\downarrow}\rangle \\
\langle\varphi|\psi_{\downarrow\uparrow}\rangle \\
\langle\varphi|\psi_{\downarrow\downarrow}\rangle
\end{matrix} \right) \;.
\end{equation}
The spin matrix operators in Eq.~(\ref{spin1}) have the form
\begin{subequations}
\label{smatrix}
\begin{equation}
\sigma^L_x =
\left( \begin{array}{cc}
0 & \mathbbm{1}  \\
\mathbbm{1} & 0  \\
\end{array} \right) \;,
\quad \sigma^L_y = i
\left( \begin{array}{cc}
0 &  - \mathbbm{1}  \\
\mathbbm{1} & 0  \\
\end{array} \right) \;,
\quad \sigma^L_z =
\left( \begin{array}{cc}
\mathbbm{1} & 0  \\
0 & -\mathbbm{1}  \\
\end{array} \right) \;,
\end{equation}

\begin{equation}
\sigma^R_x =
\left(\begin{array}{cc}
\sigma_x & 0  \\
0 & \sigma_x  \\
\end{array} \right) \;,
\quad \sigma^R_y =
\left( \begin{array}{cc}
\sigma_y & 0  \\
0 & \sigma_y  \\
\end{array} \right) \;,
\quad \sigma^R_z =
\left( \begin{array}{cc}
\sigma_z & 0  \\
0 & \sigma_z  \\
\end{array} \right) \;.
\end{equation}\\
\end{subequations}
The operator of the $k$th component of the total spin is defined as
$S^{tot}_k=(\hbar/2)(\sigma^{L}_k+\sigma^{R}_k)$.

\section{Setting up the computer experiment}

We simulate the process of swapping the spins for two electrons
localized in two QDs, which are separated by the potential
barrier.
We consider both the vertical and lateral geometry of the QD nanostructure.
In the process studied, the spin of each electron localized
in the left or right QD is flipped.
The spins are swapped as a result of the controlled
switching of the exchange interaction between the electrons.
We discuss two methods to control the exchange interaction.
According to the first method, the exchange interaction
is switched on (off) by lowering (rising) the potential barrier.
The second method is based on the changes of the depth of one of
the potential wells, which leads to the flow of electrons and their
localization in the same QD.  In the QD-based nanodevices,
both the methods can be implemented by changing
the external voltages, which are the source of the electrostatic
field forming the coupled QDs \cite{Engel2004}, or by locating the QD system
in the electromagnetic field \cite{Levy2001}.

Based on the analogy with Rabi oscillations in a two-level quantum system,
the process of switching on/off the exchange interaction that leads to
the change of the initial spin orientation to the opposite one,
will be called the $\pi$ pulse.
This process lasts for time $T^{\pi}$.
We shall also deal with processes, after which the expectation value
of $z$ electron spin component does not reach
$\pm \hbar/2$, i.e., $|\langle S_z^j \rangle| < \hbar/2$.
In such processes, the change of spin orientation is not complete.
In electrostatic QDs, the $\pi$ pulse can be realized by the proper change of
gate voltages.

The additional purpose of the present study is the optimization
of nanodevice parameters and time changes
of the confinement potential
in order to make the spin interchange time
possibly short.  The minimization of $T^{\pi}$ is important
since this time interval determines the duration of elementary quantum logic
gate operation that should be much shorter than the coherence time
of qubits.  It is required \cite{DiVincenzo1998} that the ratio
of the gate operation time to the coherence time should be less than $10^{-4}$
in order to complete a computation and several error correction runs \cite{Preskill1998}
before the decoherence destroys the information stored in qubits.
In GaAs, the spin coherence time has been estimated to be longer than 1 ms
\cite{Elzerman2004}.

At the initial time instant in each simulated process,
the system is always in the singlet ground state,
in which the electron localized in the left (right) QD
possesses the spin $z$ component $S^L_z (t=0)=+\hbar/2$
($S^R_z(t=0) = -\hbar/2$).
The two-electron wave function that satisfies
these initial conditions has a form
\begin{equation}
\label{Psi_initial}
\Psi_i\equiv\Psi(z_1,\sigma_1,z_2,\sigma_2,t=0)
= \left(\begin{matrix} 0 \\
\quad\varphi(z_1,z_2) \\
-\varphi(z_2,z_1) \\ 0
\end{matrix} \right) \;,
\end{equation}
where $\varphi(z_1,z_2)$ is the reference wave function defined in
Sec. II.
Wave function (\ref{Psi_initial}) fulfills  symmetry
constraints (\ref{psi_cond}).
After a normalization, the wave function (\ref{Psi_initial})
is subjected to a temporal evolution
according to the time-dependent Sch\H{o}dinger equation
\begin{equation}
\label{Schrtime}
i\hbar \frac{\partial \Psi(z_{1},z_{2},t)}{\partial t}
= H \Psi(z_{1},z_{2},t) \;,
\end{equation}
where $H$ is given by Eq. (\ref{Hamil}) and the time dependence
of potentials $U(z_1,z_2,t)$ and $U_{e\!f\!f}(z_1,z_2,t)$
will be determined in the following sections.

The spin expectation values are calculated using
Eq.~(\ref{spin1}) as follows.  First, the reference wave function
$\varphi(z_1,z_2)$ is calculated by the imaginary time
step method \cite{ITS}.  The reference wave function is next
applied to construct total wave function $\Psi_i$
(\ref{Psi_initial}) for $t=0$.  Wave function $\Psi_i$
is used as the initial condition to perform
the time simulation of Schr\H{o}dinger equation (\ref{Schrtime}).
In the last step, we apply $\varphi(z_1,z_2)$ to
calculate $u$ (Eq.~(\ref{u}) and $S_k^l$  (Eq.~(\ref{spin1}).

The results of simulations are presented as a function
of duration time $T$ of the process of switching
the exchange interaction.  We note that we are dealing with
two time intervals $T$ and $T^{\pi}$, which have a different meaning.
Time interval $T$ determines the duration of the arbitrary process of
changing the exchange interaction.
Time interval $T^{\pi}$ determines the duration of such process,
during which the spins are fully swapped.
At the end of each simulation, we record the expectation values
of $z$ spin component of the electron in the left (L) and right (R) QD.
These final values of spins are
denoted by $S_f^{L,R}$, where $S_f^{L,R} = S^{L,R}(t=T)$.
Moreover,  we record the energy difference $\Delta E_{fi}=E_f-E_i$
between the energies of the final ($E_{f}$) and initial ($E_{i}$) states
of the system.  The non-zero value of $\Delta E_{fi}=E_f-E_i$
gives the evidence that the process in non-adiabatic.

The numerical procedure applied allows us to observe the spins
of single electrons in another manner.  Namely, we can directly
determine the temporal evolution of the second ($\psi_{\uparrow\downarrow}$)
and third ($\psi_{\downarrow\uparrow}$) component of the
total wave function (\ref{Psivec}), whereas -- due to the symmetry --
it is sufficient to detect one of these components.
The full swapping of spins corresponds to the
interchange of these two components, i.e.,
\begin{equation}
\label{flip}
\left(\begin{matrix}
0\\
\psi_{\uparrow\downarrow} \\
\psi_{\downarrow\uparrow} \\
0
\end{matrix} \right)
\qquad\longleftrightarrow\qquad
\left(\begin{matrix}
0\\
\psi_{\downarrow\uparrow} \\
\psi_{\uparrow\downarrow} \\
0
\end{matrix} \right) \;.
\end{equation}

\section{Vertically coupled quantum dots}

The vertically coupled QDs \cite{Austing2004} can be described by the simple
model with two rectangular potential wells separated by the
rectangular barrier (Fig. 1).
The vertical QDs were fabricated on the basis of GaAs/InGaAs/AlGaAs
heterostructure \cite{Austing2004,Tarucha1996}.
The exchange interaction is switched on
by lowering the height of potential barrier to the potential
well bottom.  The exchange interaction is switched off
by raising the potential barrier to the initial height.
The time dependence of the process of changing the barrier
height can chosen in many ways.  In the present paper,
we have chosen the time dependencies described by
the step, linear, and smooth functions.

We choose such values of the parameters of the nanostructure
(Fig. 1) that the electron tunneling is negligibly small
in the initial and final time instant.
Then the electrons are well separated in space and
localized in the left (right) QD.
Owing to this the spin states of both electrons
can be exactly known in both the initial
and final instant.
We choose the same widths of the potential wells and potential barrier
 $d_{L}=d_{R}=d_B=10$ nm and
the potential well depth $V_{L}=V_{R}=-150$ meV.
At the initial time instant, we take on
the top barrier energy $V_B=0$ (the conduction band
bottom of the barrier material is chosen as the reference
energy and set equal to zero).

\begin{figure}[!ht]
\begin{center}
\includegraphics[width=0.45\textwidth]{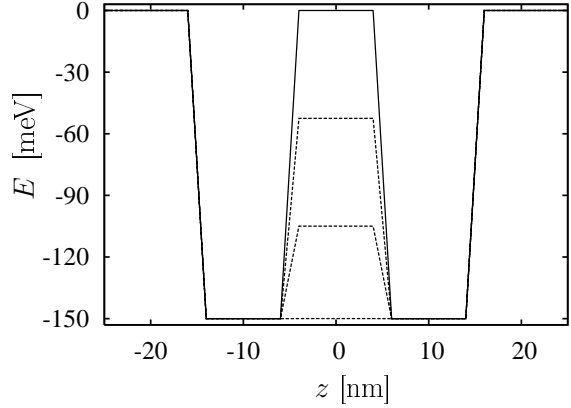}
\caption{Confinement potential profile $E\equiv V$ as a function
of vertical coordinate $z$ and energy $V_B$ of the top of
potential barrier. (a) $V_B=0$, (b) $V_B=-50$ meV,
(c) $V_B=-100$ meV, and (d) $V_B=-150$ meV.}
\label{fig1}
\end{center}
\end{figure}

\subsection{Time changes of the potential barrier}

First, we study the effect of the time step dependence of the barrier height
[Fig. 2(a)].  In this process, the top of the barrier
is rapidly lowered from $V_B=$ to $V_B=-150$ meV, i.e.,
the potential well bottom.  We let the system to evolve
for time $t=T$, after which the barrier is rapidly
raised to the starting value. We have performed
a series of simulations for different operation time $T$.
At the beginning and the end of each process, we have
determined $z$ spin component expectation values
$S_{i,f}^{L,R}$.

\begin{figure}[!ht]
\begin{center}
\hspace{
0.03\textwidth} (a) \hspace{0.3\textwidth} (b) \hspace{0.3\textwidth} (c)\\
\includegraphics[width=0.3\textwidth]{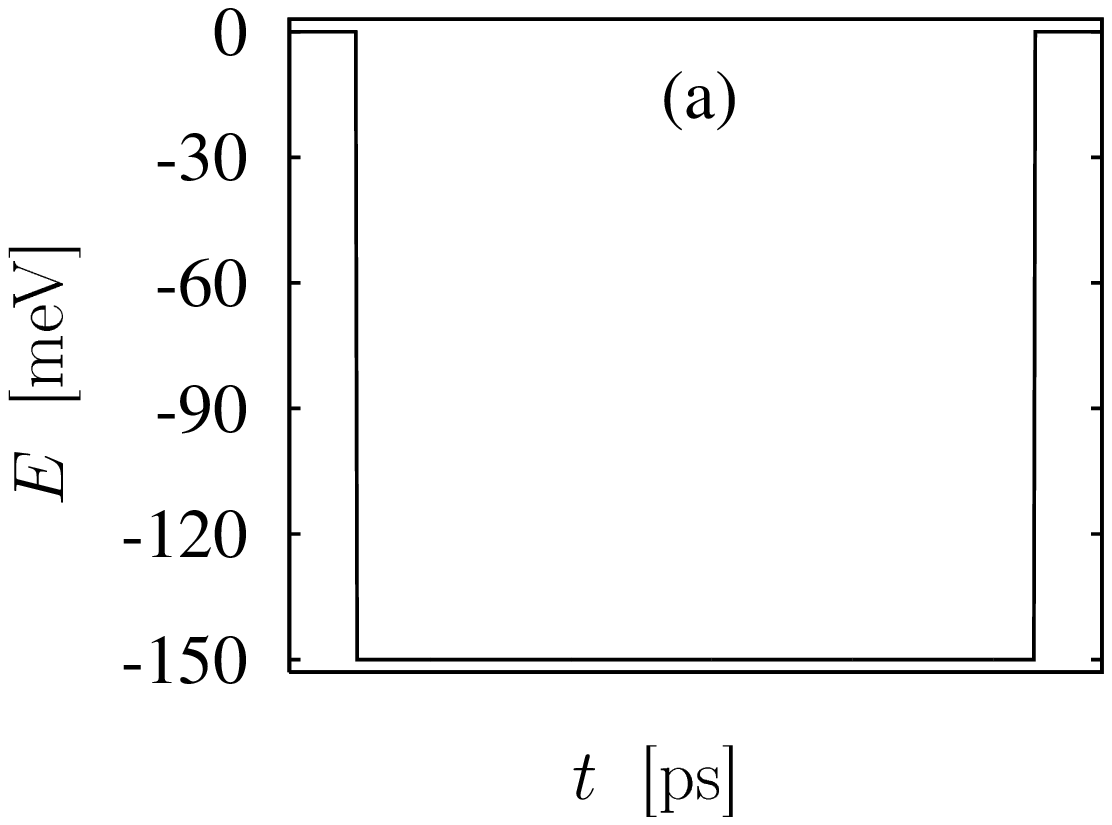}\hspace{0.03\textwidth}
\includegraphics[width=0.3\textwidth]{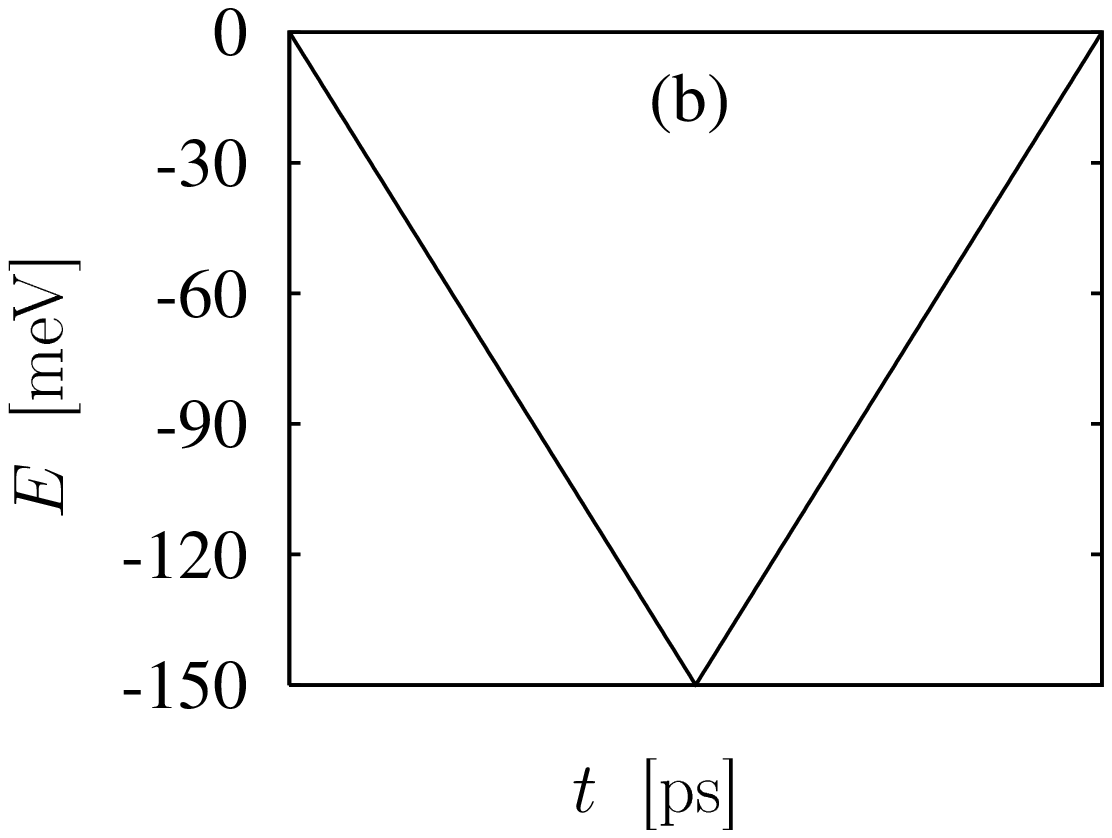}\hspace{0.03\textwidth}
\includegraphics[width=0.3\textwidth]{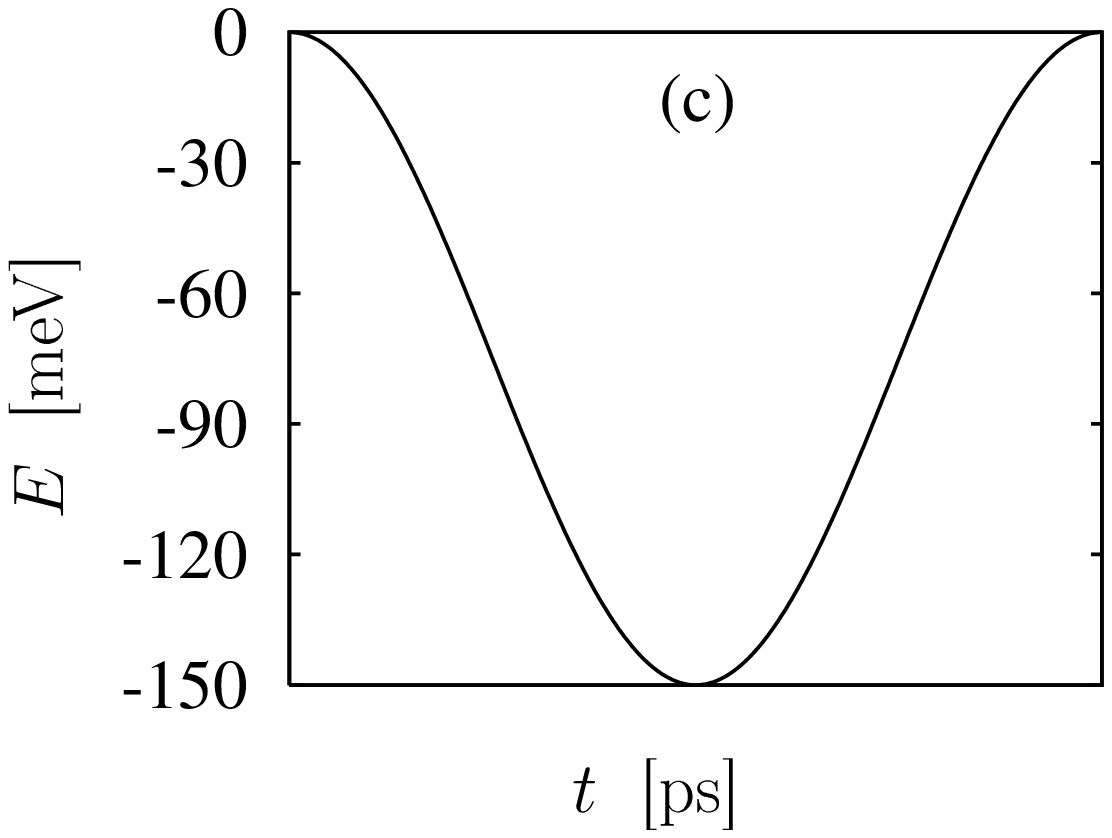}
\caption{Time dependence of top barrier $E\equiv V_B$ changes:
(a) step function, (b) linear function, (c) smooth (cosinus) function.}
\label{fig2}
\end{center}
\end{figure}

The results, displayed in in Fig. 3(a), show that the spin expectation
value detected at the final instant of each process
rapidly oscillates as a function of time duration $T$ of the process.
However, the amplitude of these oscillations does not reach $\hbar/2$.
Therefore, the sudden changes of the barrier do not lead to
to the full interchange of spins at any time interval $T$ studied.
The behaviour of energy difference $\Delta E_{fi}$ between the final and
initial state [Fig. 3(b)] allows us to explain this effect.
Similar to the average electron spin, energy difference $\Delta E_{fi}$
exhibits rapid oscillations as a function of duration
time $T$ of the process.  We interpret the results of Fig. 3(b) as follows:
during the sudden jumps of the barrier height,
the electron system makes transitions to the excited states.
As a consequence, the energy of electrons in the final state increases,
i.e., the energy difference is always larger than zero
and exhibits jumps,
and the electron wave function spreads out over the two potential
wells.  Therefore,  the switching of the interaction between the electrons
in a rapid manner is not an adiabatic process.

\begin{figure}[!ht]
\begin{center}
\hspace{0.05\textwidth} (a) \hspace{0.5\textwidth} (b) \\
\includegraphics[width=0.45\textwidth]{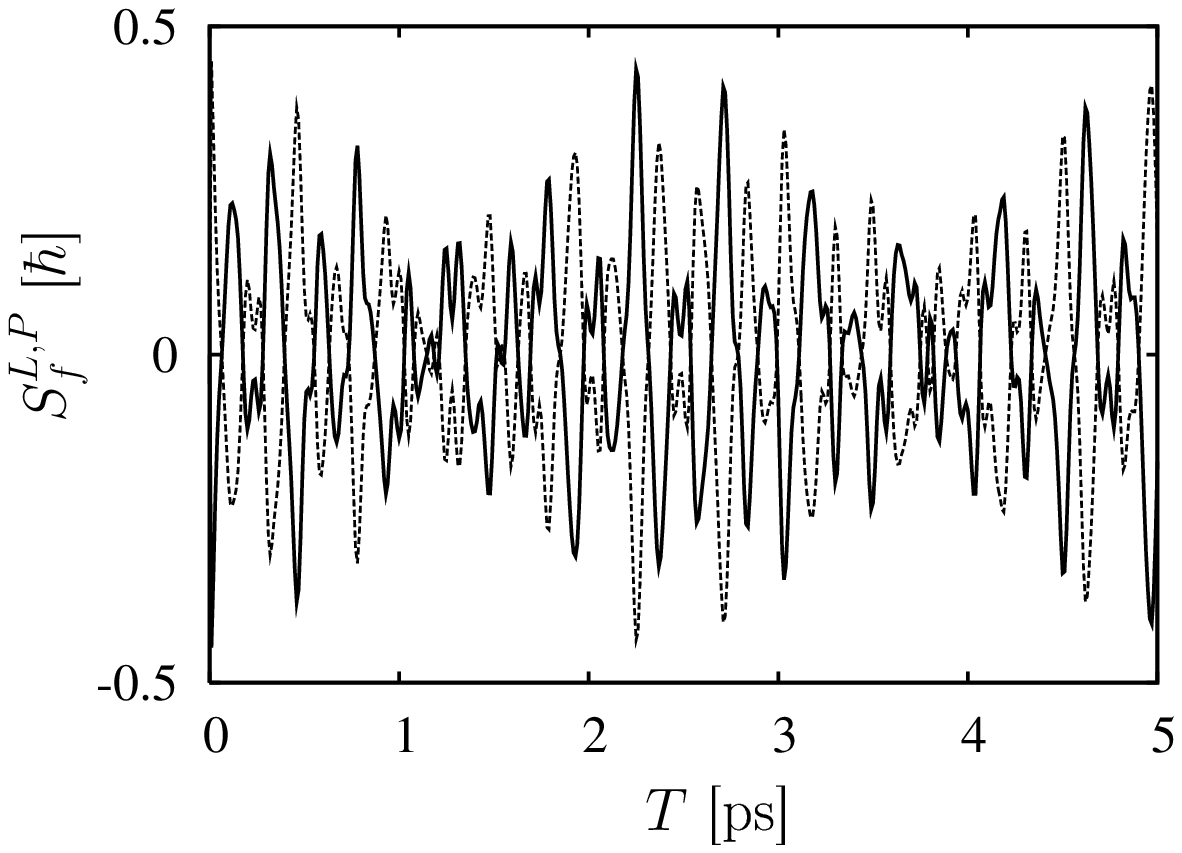}\hspace{0.08\textwidth}
\includegraphics[width=0.45\textwidth]{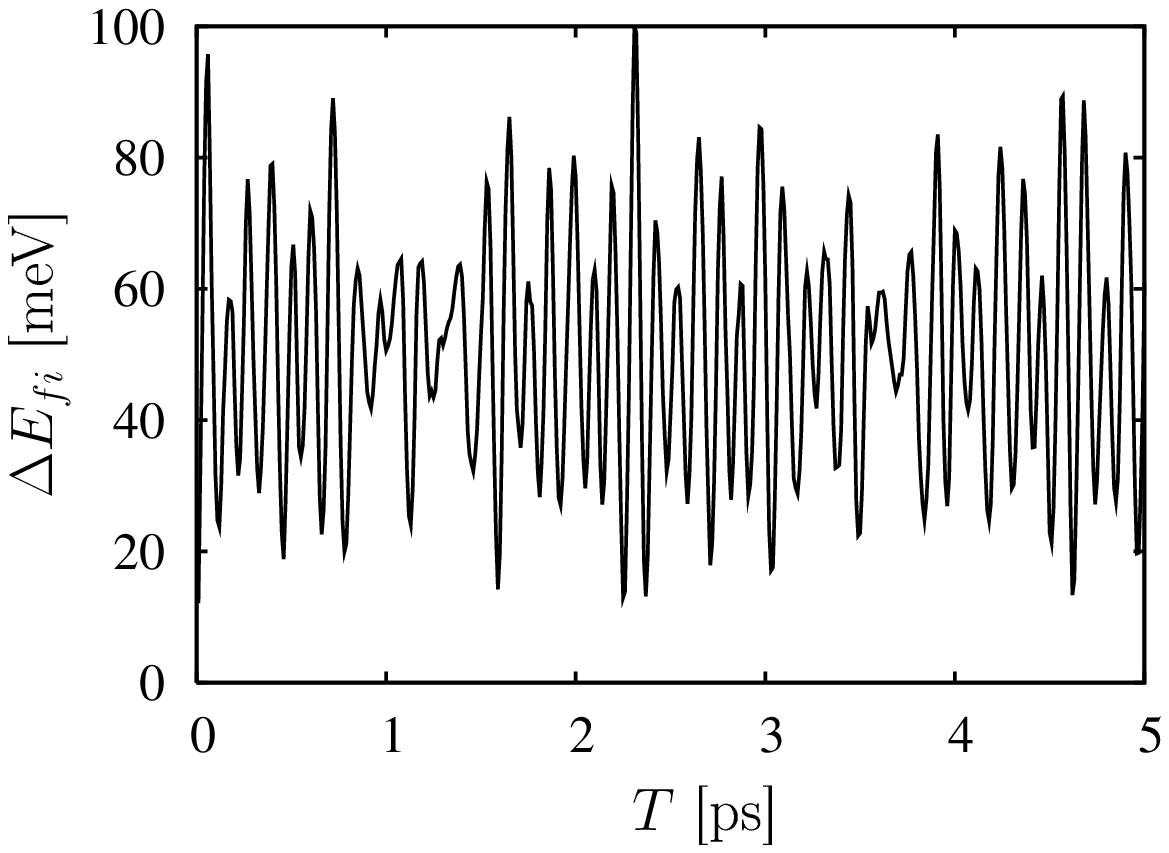}
\caption{(a) Expectation value $S_f^{L,P}\equiv S_f^{L,R}$ of $z$ spin component detected at the
end of the process of switching on and off the exchange interaction as a function
of its duration time $T$ for step-like changes of the barrier.
Solid (dashed) curve shows the results for the electron
in the right (left) QD.
(b) Energy difference $\Delta E_{fi}$ between the final and initial
states energies as a function of duration time $T$ of the process.}
\label{fig3}
\end{center}
\end{figure}

Next, we study the two-step process, during which the potential barrier
is a linear function of time [Fig. 2(b)].
In the first step, for $0\leq t \leq T/2$,
the top barrier energy is lowered according to
\begin{equation}
\label{step1}
V_B(t)=V_B^{max}-2(V_B^{max}-V_B^{min})\frac{t}{T} \;,
\end{equation}
where $V_B^{max}$ ($V_B^{min}$) denotes the largest (smallest) energy
of the barrier top in time interval $T$.
In the second step, for $T/2 \leq t \leq T$, the potential barrier is raised
according to
\begin{equation}
\label{step2}
V_B(t)=V_B^{min}+2(V_B^{max}-V_B^{min})\left(\frac{t}{T}-\frac{1}{2}\right) \;.
\end{equation}
The results (Fig. 4) show that spin expectation values
$S_{f}^{L,R}$ are periodic functions of time $T$.
Each maximal (minimal) value $S_f^R=+\hbar/2$
($S_f^L=-\hbar/2$) of $z$ spin component
corresponds to the state, in which both the electrons
possess the spins of the opposite orientation with respect
to that in the initial state.
Therefore, there exists a series of duration times $T$, for which
the $z$ spin component of each electron changes its sign,
i.e., the electrons fully exchange their spins.
The subsequent extrema in Fig. 4(a) correspond to
the $\pi$ pulses with duration time $T^{\pi}_n=n T^{\pi}_1+const$,
where index $n$ numerates the extrema.
If we increase process duration time $T$,
we simultaneously increase the time interval during which
both the electrons are localized in the same region of space.
As a consequence, for the sufficiently long time $T$
the spin swapping can occur many times.  If the process
lasted for infinitely long time, the spin interchange
would occur infinitely many times.  This scenario would be
realized for an isolated system (non-interacting
with an environment).  In this case, spin expectation
values $S_f^{L,R}$ would oscillate with amplitude
$\hbar/2$ infinitely many times, i.e., there exists
an infinite series of time intervals $T^{\pi}_n$,
for which the spins are fully interchanged.
However, in real systems, we deal with
the decay of quantum states, which leads to the energy
dissipation,  and the decoherence, which randomly changes the relative
phase of spin qubits.  These processes will lead to the
decreasing amplitude of $S_f^{L,R}$ for increasing
time interval $T$.  Therefore,
in order to perform successful quantum logic operation
the spin swapping process should last for possibly short time.
However, there exists the lower limit on time interval $T$,
during which the barrier is changed.
This limit results from the requirement of the adiabaticity
of the process: the barrier has to be changed sufficiently
slowly, i.e., in an adiabatic manner, in order to leave
the system in the ground state.  If time interval $T$
is too short, the system can be excited to higher-energy excited states,
which leads to the increase of its energy
and the non-adiabaticity of the process.

The non-adiabaticity of the spin interchange process can be also observed
in the behaviour of $S_f^{L,R}(T)$ [Fig. 4(a)].
A closer look at the present computational results
has led us to the observation that the first maximum
is slightly lower than the other maxima.
For the first maximum the exact spin
expectation value reaches $\hbar/2$ with accuracy
98.7\%.  Similarly, due to the symmetry,
for the first minimum, $S_f^L$ reaches the value $-\hbar/2$
with the same accuracy.  The extrema, which correspond to
the second $\pi$ pulse, reach $\pm\hbar/2$ with accuracy
99.4\%.  The next extrema tend to $\pm\hbar/2$
with a high accuracy.  Therefore,  we deal
with several processes of incomplete interchange of spins
before $S_f^{L,R}$ start to switch between $\pm\hbar/2$.
The lowering of the amplitude of $S_f^{L,R}$ is more
pronounced if the process of changing the potential barrier
is more rapid.  The present results show that
several processes of incomplete spin swapping
can occur before we achieve the full interchange of spins,
which leads to an elongation of the spin exchange time.
We note that this effect is not expected within the model,
based on the effective Heisenberg Hamiltonian.
In this model, the exchange interaction between
the electrons is described by the Hamiltonian
$H(t) = J(t)\mathbf{S}_1 \cdot \mathbf{S}_2$, where $J(t)$
is exchange interaction energy and
the spin operators $\mathbf{S}_j$ ($j=1,2$) are
expressed in terms of the corresponding Pauli matrices.
Our estimates show that -- in real quantum systems --
the spin interchange time will be longer than that
resulting from the Heisenberg model, which do not take
into account the physical implementation of switching on/off
the exchange interaction.

\begin{figure}[!ht]
\begin{center}
\hspace{0.05\textwidth} (a) \hspace{0.5\textwidth} (b) \\
\includegraphics[width=0.45\textwidth]{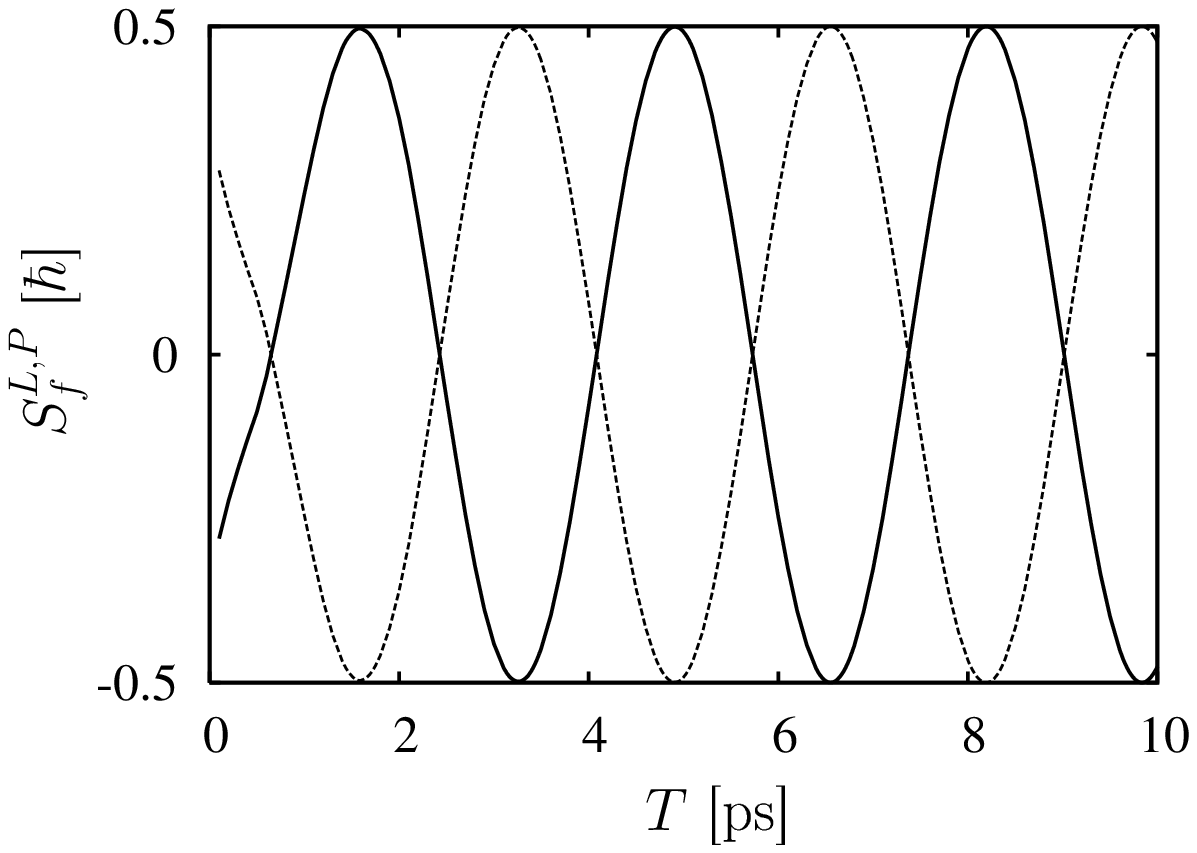}\hspace{0.08\textwidth}
\includegraphics[width=0.45\textwidth]{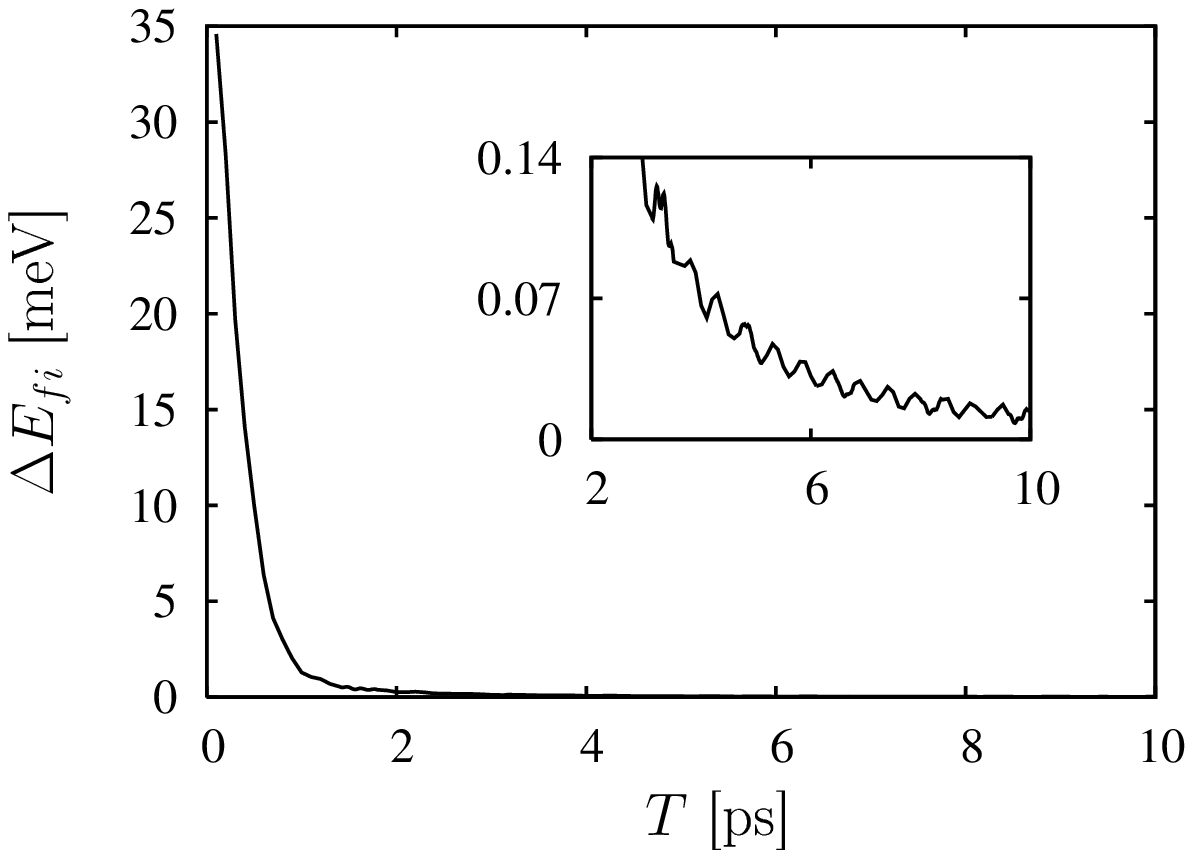}
\caption{(a) Expectation value $S_f^{L,P}\equiv S_f^{L,R}$ of $z$ spin component detected at the
end of the process of switching on and off the exchange interaction as a function
of duration time $T$ for linear changes of the barrier.
Solid (dashed) curve shows the results for the electron
in the right (left) QD.
(b) Energy difference $\Delta E_{fi}$ between the final and initial
state energies as a function of duration time $T$ of the process.
Inset zooms in the part of this figure.}
\label{fig4}
\end{center}
\end{figure}

Fig. 4(b) displays the dependence of energy separation $\Delta E_{fi}$
on time interval $T$.  In the inset, we zoom in this dependence
for the narrower scale of $T$, which allows us to observe the oscillations
of $\Delta E_{fi}$.  The amplitude of these oscillations
decreases if the duration time of spin interchange increases.
Moreover, $\Delta E_{fi}$ asymptotically approaches zero
for large $T$.  For small $T$ the positions of the local minima of $\Delta E_{fi}$
do not coincide with $T^{\pi}_n$ [cf. Fig. 4(a)].  For the chosen nanostructure parameters,
this coincidence appears for the third $\pi$ pulse at
$T=8.2$ ps.
For linear changes of the barrier we deal with rapid turning on
the process at $T/2$, which causes the non-adiabaticity of this method
of spin interchange.

Based on the above results, we expect that we will obtain better results if
the temporal changes of the potential barrier will be modelled
by a smooth function of time.  In the following, we study
the smooth time dependence of the potential barrier, which
is modelled by the following function [Fig. 2(c)]:
\begin{equation}
\label{cos}
V_B(t)= \frac{1}{2}(V_B^{max}-V_B^{min}) (1-\cos\omega t) + V_B^{max} \;,
\end{equation}
where $\omega=2\pi/T$.  The results of simulations
[Fig. 5(a)] show that the amplitude of $S_f^{L,R}(T)$ approaches $\hbar/2$
much faster than for step-like and linear changes of the barrier.
Already the second maximum reaches $\hbar/2$ with accuracy 99.99\%.
Moreover, based on the results for energy difference $\Delta E_{fi}$
[Fig. 5(b)] we can regard this process to be
adiabatic for $T\geq 2$ ps.  The very fine jumps of $\Delta E_{fi}$
near zero, shown in the inset of Fig. 5(b), result from the
small numerical errors.  Therefore, the amplitude of the periodic
changes of $S_f^{L,R}$ remains constant (and equal to $\hbar/2$) for $T > 2$ ps.

\begin{figure}[!ht]
\begin{center}
\hspace{0.05\textwidth} a) \hspace{0.5\textwidth} b) \\
\includegraphics[width=0.45\textwidth]{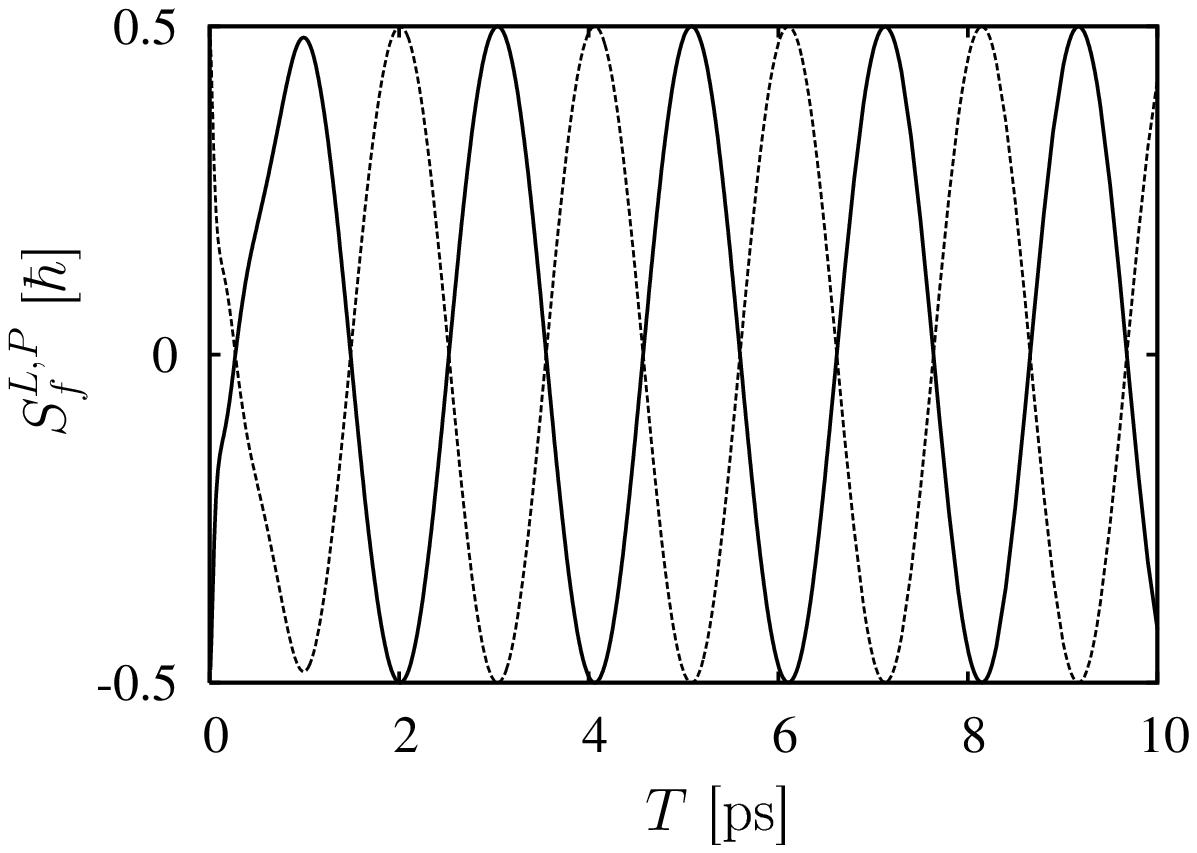}\hspace{0.08\textwidth}
\includegraphics[width=0.45\textwidth]{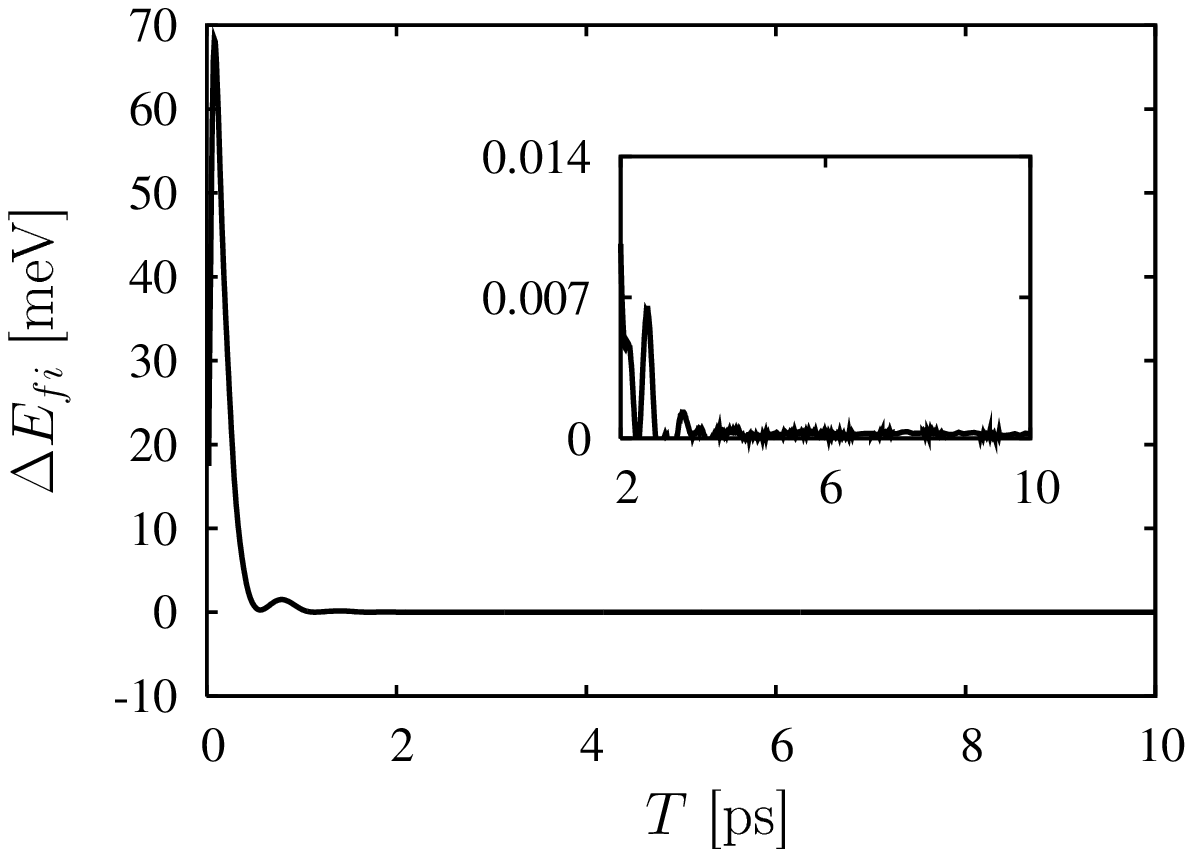}
\caption{(a) Expectation value $S_f^{L,P}\equiv S_f^{L,R}$ of $z$ spin component detected at the
end of the process of switching on and off the exchange interaction as a function
of duration time $T$ for smooth changes of the barrier.
Solid (dashed) curve shows the results for the electron
in the right (left) QD.
(b) Energy difference $\Delta E_{fi}$ between the final and initial
state energies as a function of duration time $T$ of the process.
Inset zooms in the part of this figure.}
\label{fig5}
\end{center}
\end{figure}

The spin of the electron localized in the left (right) QD can be
determined from Eq.~(\ref{spin1}) only if the probability of finding
both the electrons in the same QD is zero.  This state is realized
at the initial and final time instant in each process of the barrier changes.
At the intermediate time instants,
the electrons are not spatially separated; therefore,
quantities $S^{L,R}\equiv S^{L,R}_z$ , calculated
from Eq.~(\ref{spin1}), are not equal to the eigenvalues of $z$ spin component
of the electron in the left and right QD.
Nevertheless, for arbitrary time, $S^{L,R}$  can serve
as a control spin index.

In Fig. 6 we present the time dependence of control spin indices and
energy of the system, which correspond to the pulses with duration times
$T_1^{\pi}$ and $T_2^{\pi}$ for linear and smooth changes of the barrier.
The energy is measured relative the ground state energy $E_0$ of two electrons
in the lateral confinement potential ($E_0=2\hbar\omega_{\perp}$, where
$\hbar\omega_{\perp}=40$ meV).
The behaviour of these time characteristics is different
for the smooth and non-smooth changes of the barrier.
In the case of the non-smooth process,
the discontinuity of the first time derivative of the top barrier energy
for $t=T/2$ leads to the corresponding discontinuity
of the energy versus time plot [cf. dotted curves in Figs. 6(a) and 6(b)].
The consequences of this discontinuity can also be seen
in the time dependence of the control spin indices
[cf. solid and dashed curves in Figs. 6(a) and 6(b)].
For the pulse with time duration $T^{\pi}_1=1.58$ ps
the spin indices begin to oscillate after reversing
the time evolution of the potential barrier [Fig. 6(a)].
The amplitude of these oscillations decreases with the increasing
barrier height and falls down to zero when the barrier height
reaches the starting level.  For the three times longer
pulse $T_2^{\pi}=4.91$ ps reversing the time changes of the barrier
leads to smaller oscillations [cf. Fig. 6(b)].  Nevertheless, the time dependence
of the control spin indices is still slightly perturbed.

\begin{figure}[!ht]
\begin{center}
(a) \hspace{0.47\textwidth} (b) \\
\includegraphics[width=0.45\textwidth]{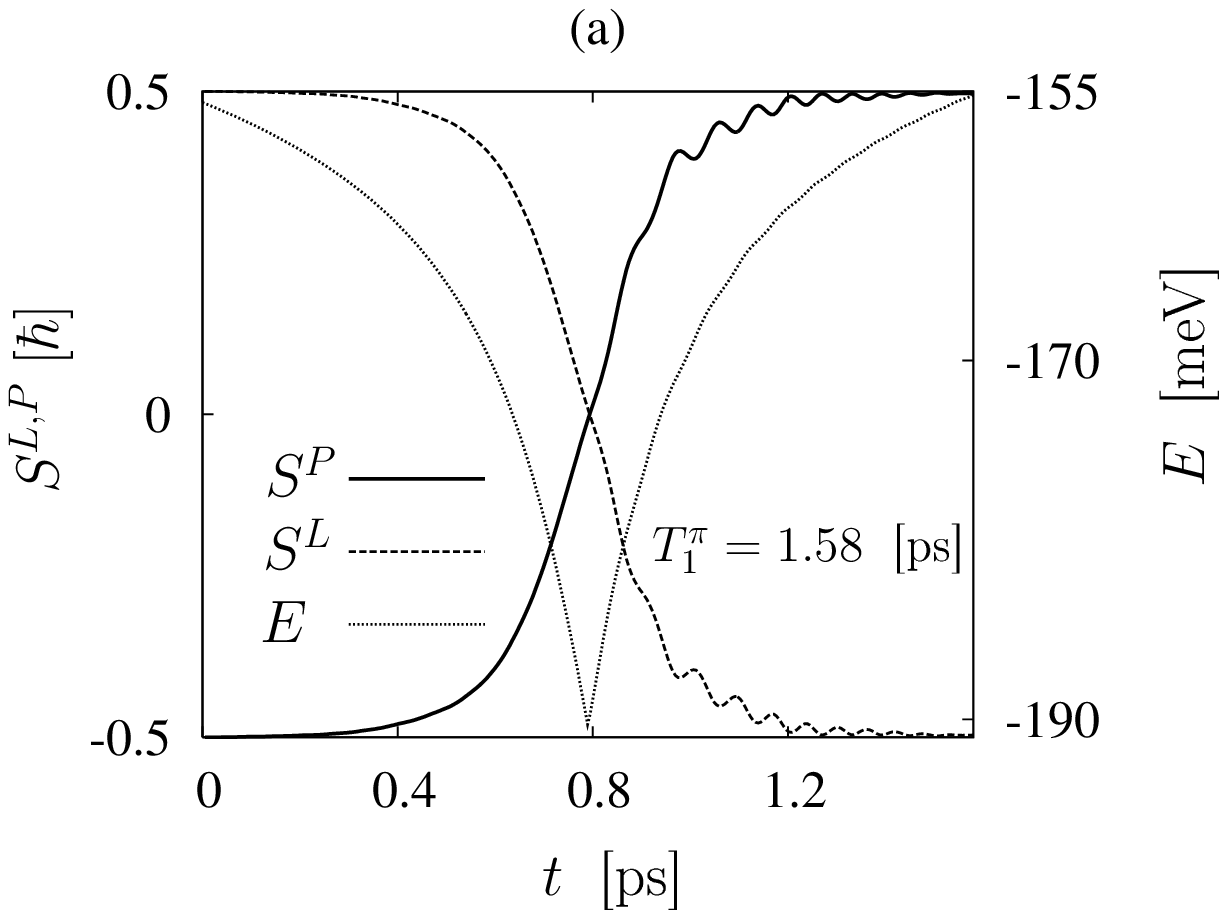}\hspace{0.08\textwidth}
\includegraphics[width=0.45\textwidth]{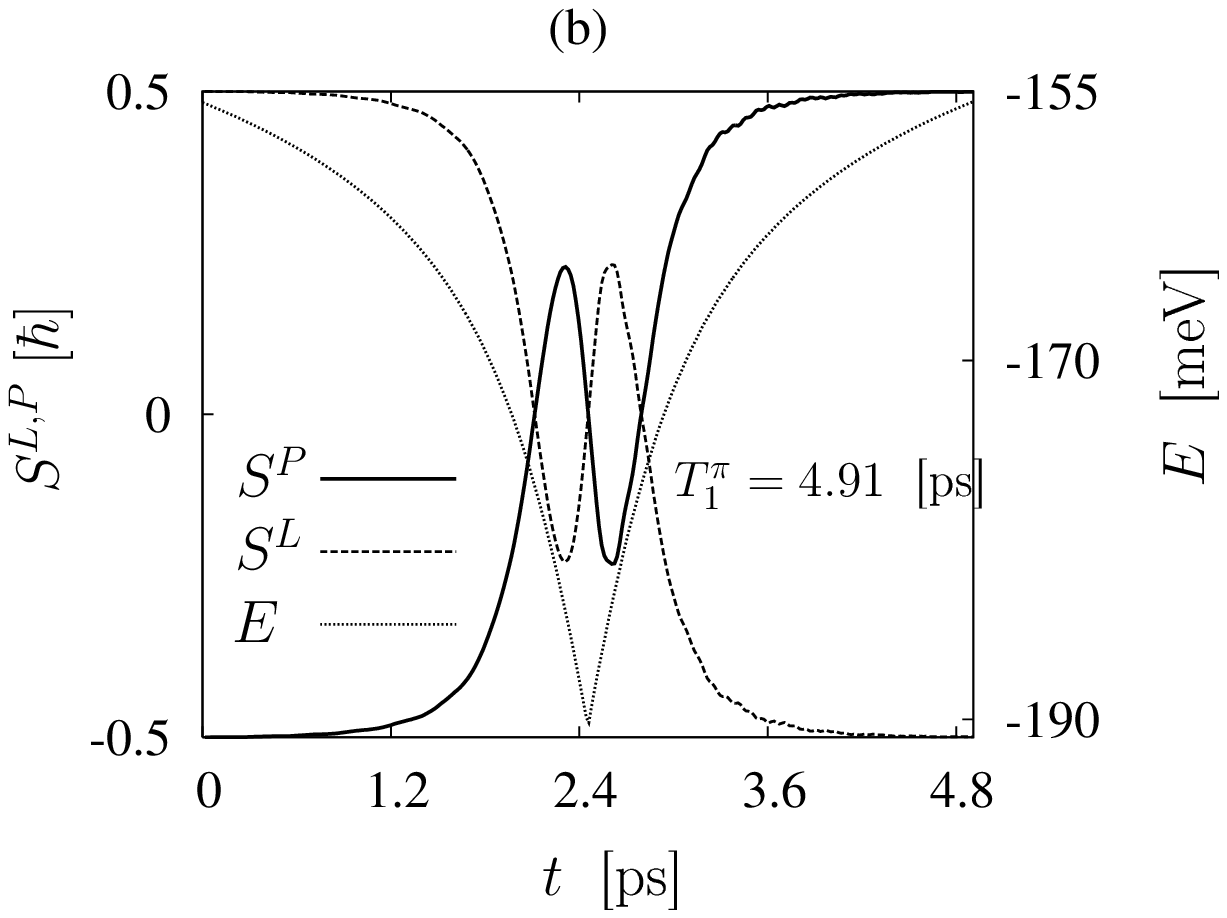}\\
\vspace{12pt}
(c) \hspace{0.47\textwidth} (d) \\
\includegraphics[width=0.45\textwidth]{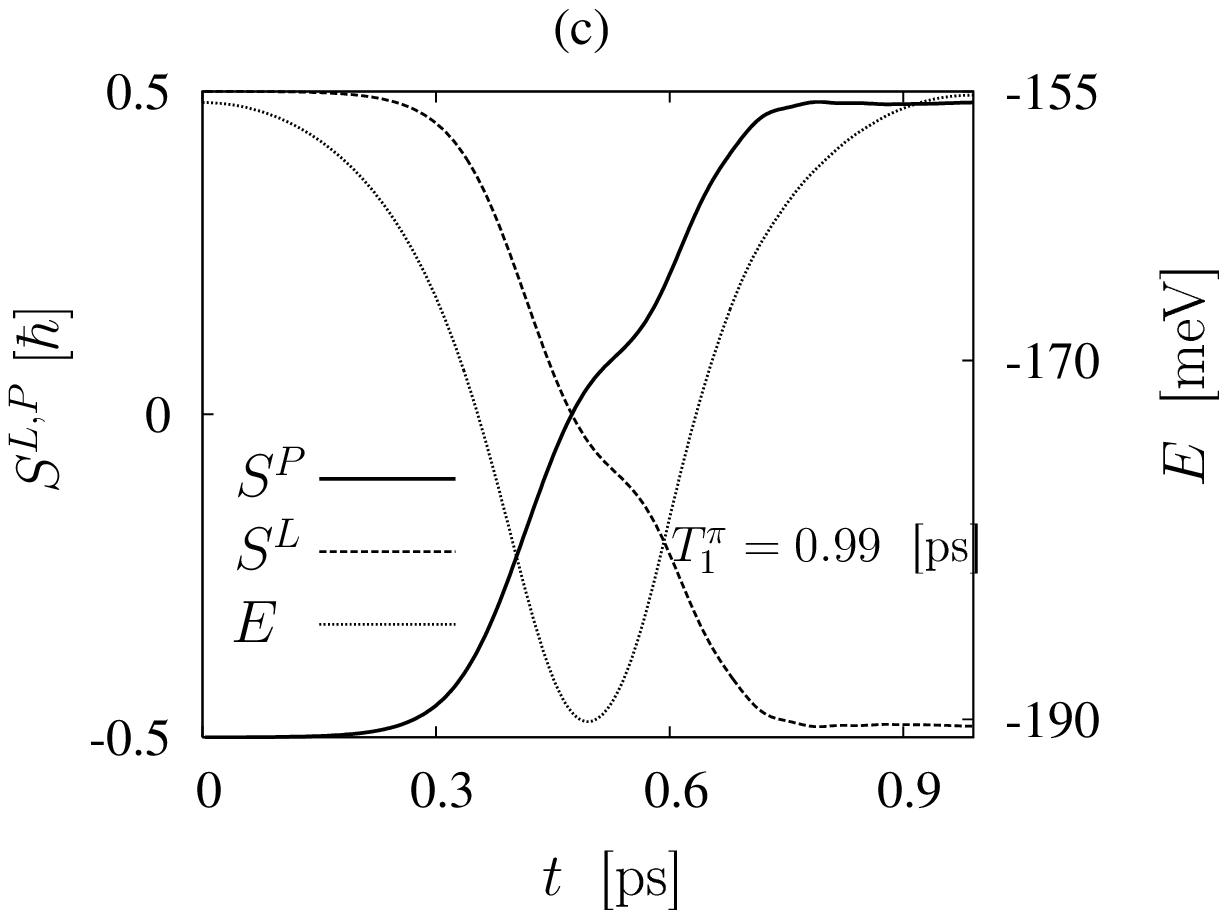}\hspace{0.08\textwidth}
\includegraphics[width=0.45\textwidth]{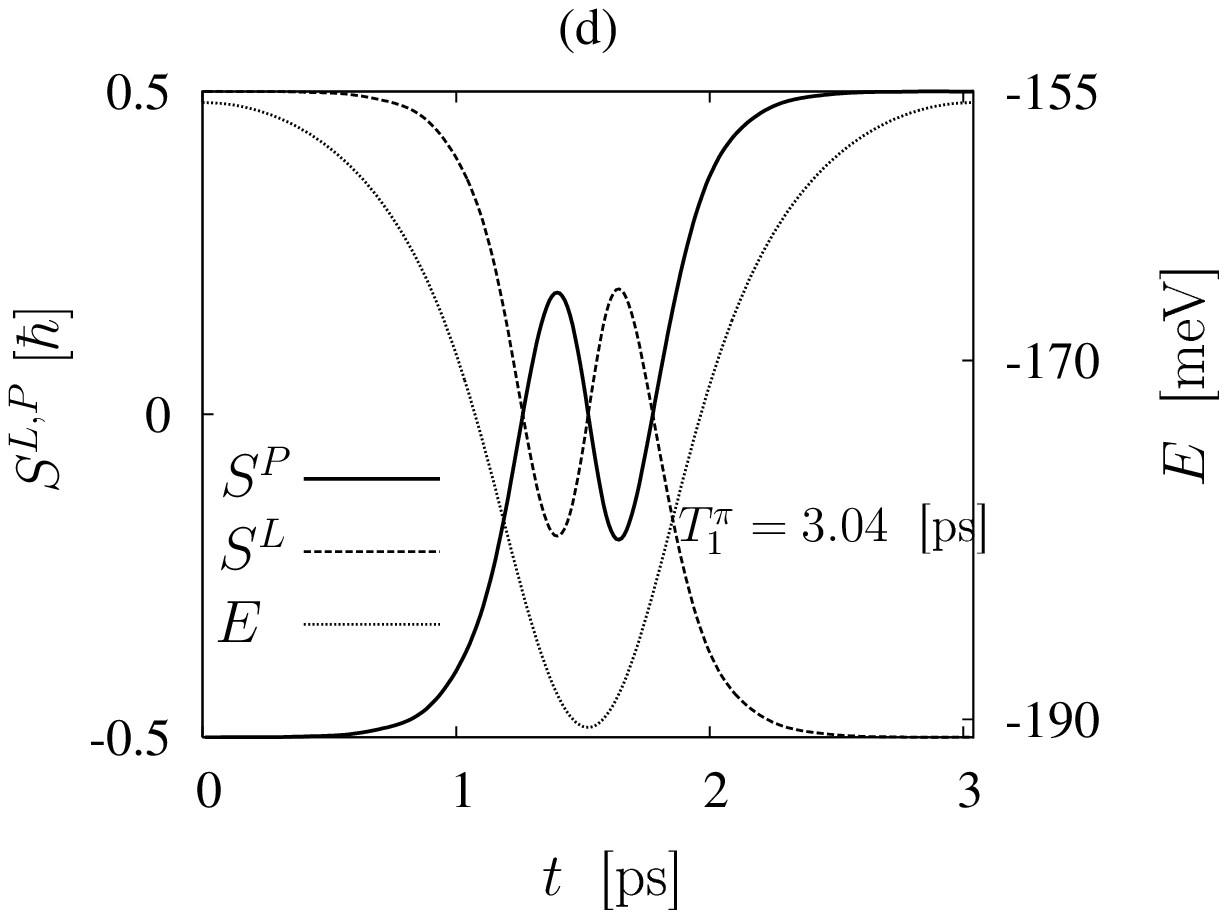}
\caption{Spin control indices  $S_f^{L,P}\equiv S^{L,R}$ and energy $E$ of the
two-electron system as functions of time $t$.
Plots (a) and (b) [(c) and (d)] correspond to the linear [cosinusoidal]
changes of the potential barrier, plots (a) and (c) [(b) and (d)]
correspond to the first [second] $\pi$ pulse with duration time
$T^{\pi}_1$ [$T^{\pi}_2$].  Solid (dashed) curves display
the control spin index of the electron in the right (left) QD
and the dotted curves show the energy of the system.}
\label{fig6}
\end{center}
\end{figure}

In the case of smooth time changes of the barrier [Eq.~(\ref{cos})],
the energy of the system is also a smooth function of time
[Figs. 6(c,d)].  In spite of this, for the first $\pi$ pulse
with $T^{\pi}_1=0.99$ ps the changes of the potential barrier
are still seen by the system as rapid changes, which leads to
bending the spin characteristics [Fig. 6(c)].
However, the time duration of the second $\pi$ pulse
($T_2^{\pi}=3.04$ ps) is sufficiently long so that
the spin characteristics [Fig. 6(d)] stay smooth during
the full process of switching on/off the exchange interaction.
In the cases of linear and smooth changes of the potential barrier,
the time evolution of spin indices $S^{L,R}(t)$ for the second
$\pi$ pulse with time duration $T^{\pi}_2$ suggests that
the full interchange of spins occurs as a result of many-fold exchange
of spins between the electrons.  For the second pulse the curves
$S^{L,R}(t)$ go through zero three times [cf. Figs. 6(b,d)].
Thus, we conclude that in these processes we are dealing
with the three-fold incomplete swapping of spins before we finally obtain
the full interchange of spin orientations.
In order to obtain an additional support for the occurrence of this process,
we have observed the time changes of the second and third
components of the total wave function [Eq.~(\ref{Psivec})].
The results of the simulations performed show the many-fold
interchange of the second and third wave function components,
which means that -- in fact -- the spins of the electrons change many times.

Based on these results, we have chosen to further studies
the method of the smooth switching of the exchange interaction,
which seems to be the most promissing in a physical realization.

\subsection{Effect of quantum dot size and potential well asymmetry}

If the size of the coupled QD nanostructure is sufficiently large,
the Coulomb interelectron repulsion leads to a localization
of electrons at the opposite potential
well boundaries \cite{Bednarek2003art}.  Then, the electrons form
the Wigner molecule \cite{Bednarek2003art}.  In the case of Wigner localization,
the lowering of the potential barrier can cause no interchange of spins
during a time, which is shorter than the spin coherence time.
We have checked the possibility of the occurrence of this effect by investigating
the influence of the QD size of the time duration of spin interchange
process.
The calculations have been performed for the QD nanostructure
modelled by rectangular potential wells and a barrier.
The barrier thickness was fixed as $d_B=10$ nm, the top barrier
energy was changed according to Eq.~(\ref{cos}),
the depths of both the potential wells were equal ($V_L=V_R=-200$ meV), and
the thicknesses of both the potential wells were also the same $d_L=d_R\equiv d$.
We change potential well width $d$ in the interval $d \in [6,50]$ nm.
The size of the region, in which the spin exchange appeared,
has the width $2d+d_B$.
Figure 7 depicts the calculated time duration of the first and second
$\pi$ pulse as a function of width $d$ of the potential wells.
We see that the time duration of spin interchange quickly increases
with increasing QD size.  The $\pi$ pulse time changes with the QD size
as follows: $T^{\pi} \sim d^3$ for
$d < 20$ nm and $T^{\pi} \sim d^6$ for $d > 20$ nm.
These results show that even for the QD with the size such small as $d \simeq 50$ nm
the full spin swapping can be not realized during the coherence time.

\begin{figure}[!ht]
\begin{center}
\includegraphics[width=0.5\textwidth]{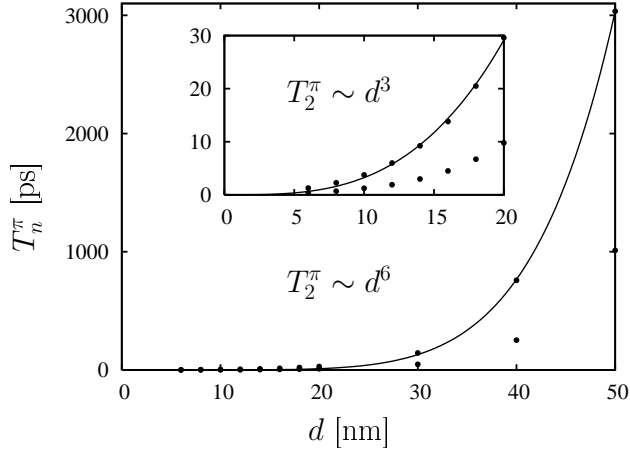}
\caption{Time duration $T^{\pi}$ of the first ($n=1$) and second ($n=2$) $\pi$ pulse
as a function of thickness $d$ of the potential well.
Inset zooms in the short time part of the figure.}
\label{fig7}
\end{center}
\end{figure}

In the Heisenberg model of electron-electron interaction,
the rate of spin interchange is determined by the exchange coupling
constant, which is defined as the energy difference
between the triplet and singlet states, i.e.,
$J=E_T-E_S$.  For the two identical QDs separated by
the sufficiently thick and high potential barrier
the triplet and singlet states are degenerate and $J=0$.
If the potential barrier becomes thinner and lower,
the wave functions of the electrons localized in both the QDs
start to overlap and the energies of triplet and singlet states become different,
i.e., $J \neq 0$.
The coupling constant $J$ can also be changed if the QD potential wells
are different, i.e., the QDs are asymmetric \cite{Szafran2004b}.  The larger the asymmetry
of the potential wells the larger coupling constant $J$.
Therefore, one could expect that the asymmetry of the QDs
should lead to a shortening of the spin interchange time.
However, it turns our \cite{Szafran2004} that this is not always
the case.

We have investigated the influence of the asymmetry of potential
well thicknesses on the duration time of spin interchange.
For the fixed thickness of the left QD ($d_L=10$ nm)
we have performed the calculations for several values of thickness
$d_R$ of the right QD.  The results are plotted in Fig. 8 for
$d_R=5$ nm and $d_R=15$ nm.
For sake of clarity, we present in Fig. 8 only the results
for the electron in the right QD.  The spin expectation values ($S_f^L$)
for the electron in the left QD can be obtained
by the reflection of $S_f^R(T)$ curves with respect
to the axis $S_f^R=0$.

The results of calculations are displayed in Fig. 8.
If the right potential well is thinner
than the left one, energy difference $\Delta E_{fi}$
between the final and initial state tends to zero
for long duration time $T$ of spin interchange [dashed curves in Fig. 8(a)].
However, if the right QD is thicker than the left one,
$\Delta E_{fi}$ does not reach zero even for
long time $T$ [Fig. 8(b)]. Moreover,
it takes on negative values for some times $T$.
The negative values of $\Delta E_{fi}$ can be explained
by the shift of electron density toward the wider QD
in which the final state energy is lower.
Analyzing the plots of $S_f^R$ vs $T$ we have observed
that the accuracy of the interchange of spins reaches
99.9 \%, but is never equal to 100 \%.
We conclude that the asymmetry of the QD confining potential
results in the increasing multiplicity of spin exchange,
which leads to the longer duration time of spin interchange.

\begin{figure}[!ht]
\begin{center}
(a) \hspace{0.47\textwidth} (b) \\
\includegraphics[width=0.45\textwidth]{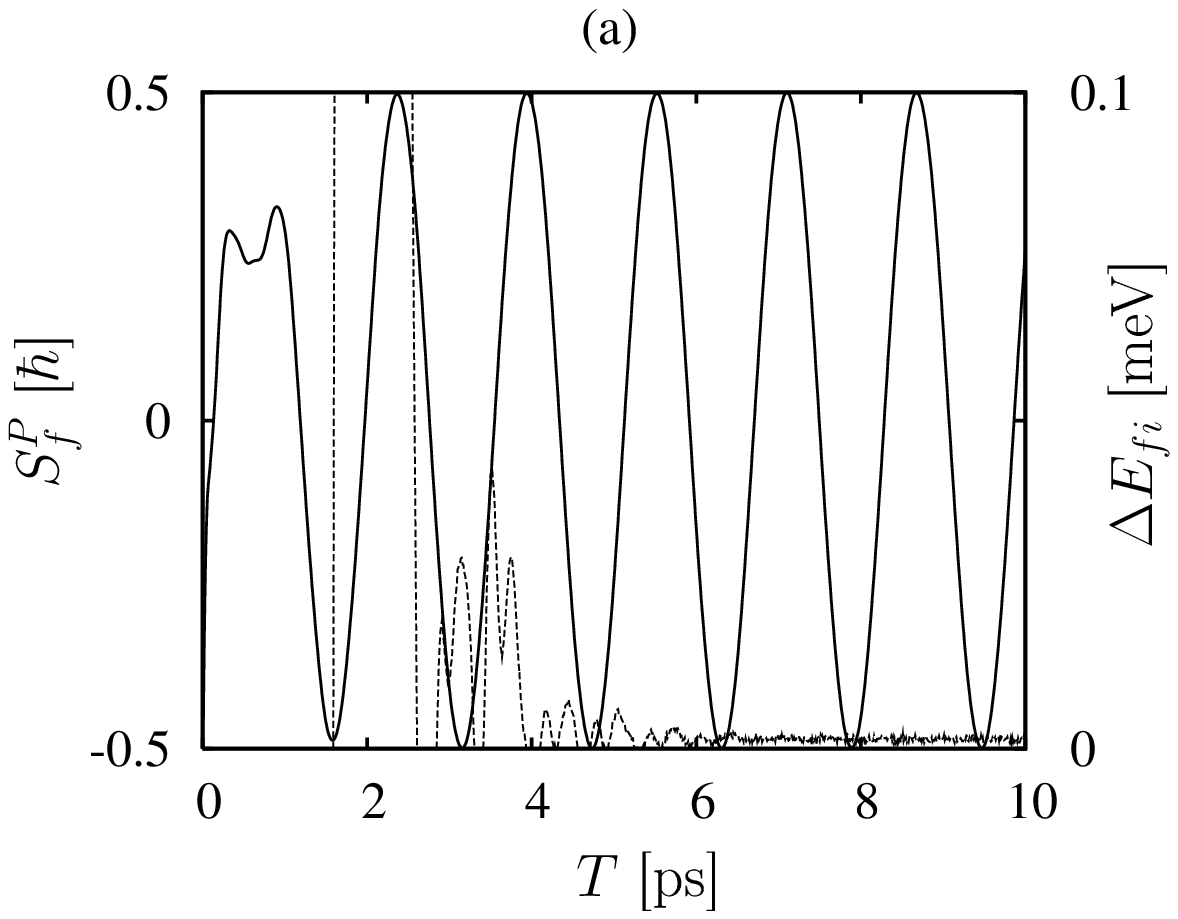}\hspace{0.08\textwidth}
\includegraphics[width=0.45\textwidth]{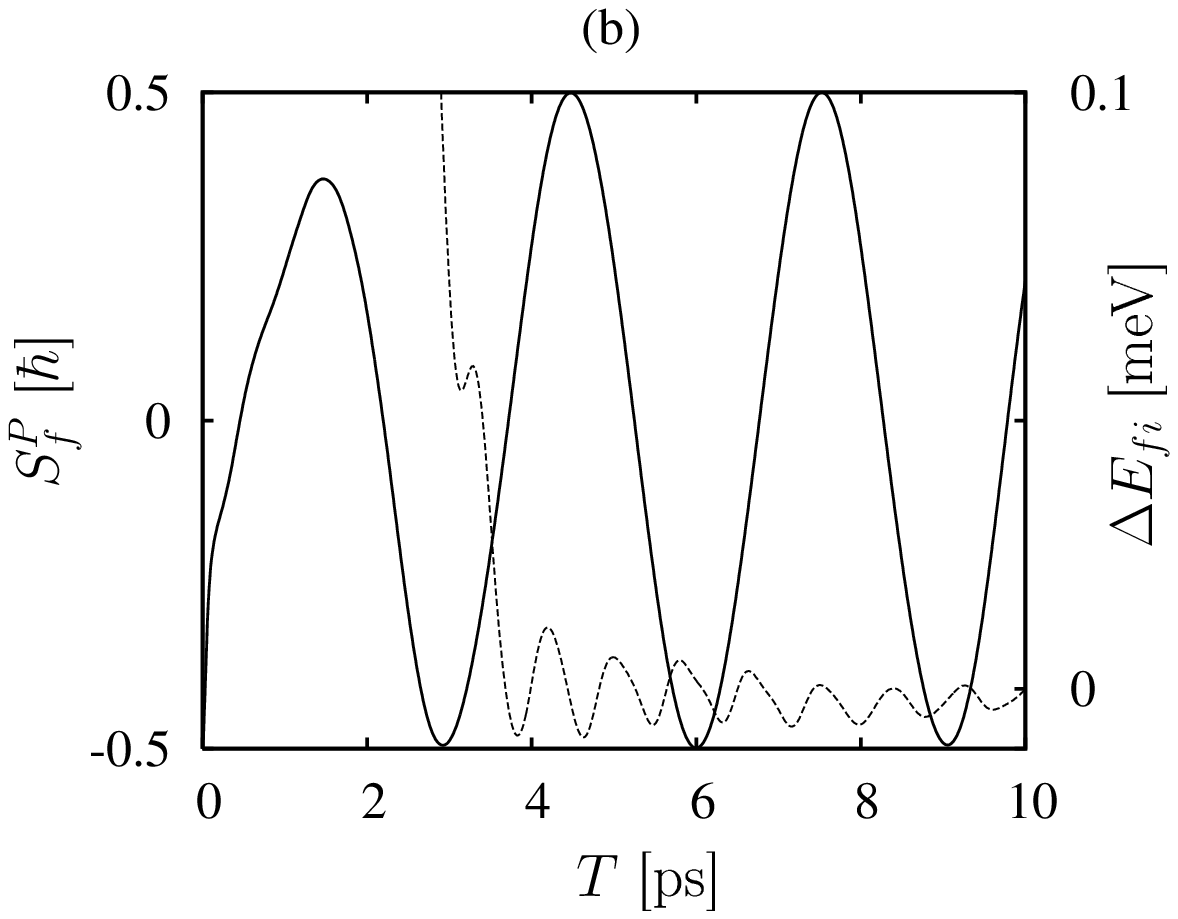}\\
\caption{Expectation value $S_f^P\equiv S_f^{R}$ of $z$ spin component detected
in the right QD  at the end of the process of switching on/off
the exchange interaction (solid curves) and energy difference
$\Delta E_{fi}$ between the final and initial states (dashed curves)
as functions of duration time $T$ of spin interchange for asymmetric QDs.
The thickness of the left QD is fixed $d_L=10$ nm,
while the thickness of the right QD $d_R=5$ nm in (a)
and  $d_R=15$ nm in (b).}
\label{fig8}
\end{center}
\end{figure}

\section{Laterally coupled quantum dots}

The electrostatic QDs with lateral interdot coupling are the
subject of many recent studies \cite{Vaart1995,Engel2001,Chang2001}.
In the laterally coupled QDs, the profile of the confinement potential
can be modified by varying the voltages applied to the gates.
This provides a fairly convenient way of tuning the potential
barriers and wells.  However, due to technological limitations,
the size of the lateral QD is of the order of
100 nm \cite{Elzerman2005}, i.e., is larger than
the size ($\sim 10$ nm) of the vertical QD \cite{Tarucha1996} measured
in the growth direction.
The nanostructure consisting of two laterally coupled QDs
separated by the potential barrier
has a typical linear size $200-300$ nm.
Due to this relatively large size, the duration time
of the spin rotation is expected to be long (cf. see Subsection IV.B).

We still apply the theoretical model described
in Section II with the frozen transverse motion
of electrons.  Now $(y,z)$ are the transverse coordinates
and we assume that the electrons move in the $x$ direction,
e.g., the $x$ axis can be directed along the effectively
one-dimensional flow of electrons between the two
lateral QDs \cite{Elzerman2005}.
The larger size of the lateral QDs in comparison
to that of vertical QDs leads to the weaker electron localization
and the weaker electron-electron interaction.
According to the theory \cite{Bednarek2003eff}
the strength of the potential confining the electrons in the transverse directions
determines the strength of the effective electron-electron
repulsion [Eq.~(\ref{Ueff})]. [We note that now $U_{e\!f\!f}$ is a function of $x$.]
The excitation energy $\hbar\omega_{\perp}$ is a measure of
the strength of the transverse parabolic confinement.
In the calculations for the lateral QDs, we take on $\hbar\omega_{\perp}=5$ meV.
Then, the transverse confinement potential rather well
approximates the confinement in the double lateral QD
fabricated by the TU Delft laboratory \cite{Elzerman2003}.
The present model is also applicable to quantum wires,
which are composed from different semiconductors, e.g.,
InP and InAs \cite{Bjork2002}, that form a structure of
potential wells and barriers.  In this case, we deal
with quasi-one-dimensional coupled QDs.
In the present section, we modify the exchange interaction
between the electrons localized in laterally coupled QDs
by tuning the potential barrier as well as the potential wells.

\subsection{Symmetric quantum dots}

The profile of the confinement potential for the laterally
coupled symmetric QDs is assumed in the form (Fig. 9)
\begin{equation}
V(x)=kx^2+V_2 \exp\left[-(x-x_0)^2/d^2\right]
\label{parGauss}
\end{equation}
where $k$ determines the strength of the parabolic confinement potential,
parameter $V_2$ is the potential barrier height for $V_2>0$
(potential well depth for $V_2<0$) , $x_0$ is the position of the center of
the potential barrier (well), and $d$ is the range of the Gaussian
potential.

The potential barrier separating the two QDs can be modified
by changing $V_2$ [Eq.~(\ref{parGauss})].
In the calculations, parameter $V_2$ was changed in time
in a smooth manner, i.e., according to Eq.~(\ref{cos}),
from 9 meV to 0 and next from 0 to 9 meV.
Figure 9 shows the corresponding profiles of the confinement potential.
We see that when changing the potential barrier we simultaneously
change the potential wells.
If the potential barrier is lowered,
the electrons become localized closer to each other.
The effective QD size ($d^{e\!f\!f}_{L,R}$) can be defined as the width
of the corresponding potential well
determined at the half of barrier height.
At the initial time moment, the effective size of the QDs
$d^{e\!f\!f}_L=d^{e\!f\!f}_R=100$ nm,
the effective width of the barrier $d^{e\!f\!f}_B= 60$ nm,
the electron in the left (right) QD has spin $+\hbar/2$ ($-\hbar/2$).

\begin{figure}[!ht]
\begin{center}
\includegraphics[width=0.45\textwidth]{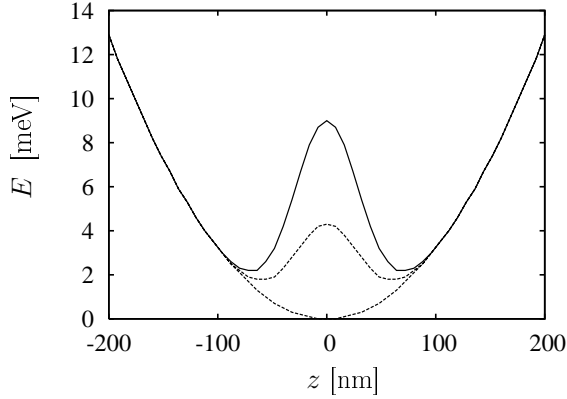}
\caption{Profiles of confinement potential $E\equiv V$ vs $z\equiv x$
for two laterally coupled QDs for $V_2 \geq 0$.
Parameters of the confinement potential [Eq.~(\ref{parGauss})] are
$k=9.0\times 10^{-7}$ meV(nm)$^{-2}$, $d=185$ nm, $x_0=0$,
$V_2=9.0, 4.3$, and  0 meV (solid, dashed, and dotted curves).}
\label{fig9}
\end{center}
\end{figure}

In Fig. 10, we display the time evolution of spin indices
$S^{L,R}$ and parameter $V_2$.
These results show that for the first $\pi$ pulse with
time duration $T_1^{\pi}=492$ ps the spin interchange occurs with 100\%
accuracy.  We see that during the larger part
of time evolution the spins do not change.
The spins start to rotate when the barrier height $V_2$
decreases to $\sim 5$ meV.
This suggests that the lowering of the potential barrier
should lead to a shorter spin rotation time.
If, however, the initial barrier height decreases,
the effective QD size decreases too.  For $V_2=5$ meV
the effective size of the potential wells and the barrier
decrease to $d^{e\!f\!f}_{L,R}= 76$ nm and $d^{e\!f\!f}_B=42$ nm.
This means that the shortening
of the spin rotation time results from the size effect,
discussed in Subsection IV.B.
The fastest rate of spin interchange is observed
for $V_2 < 1.5$ meV.  Then, the effective sizes
are $d^{e\!f\!f}_{L,R}=40$ nm and $d^{e\!f\!f}_B=42$ nm.

\begin{figure}[!ht]
\begin{center}
\includegraphics[width=0.45\textwidth]{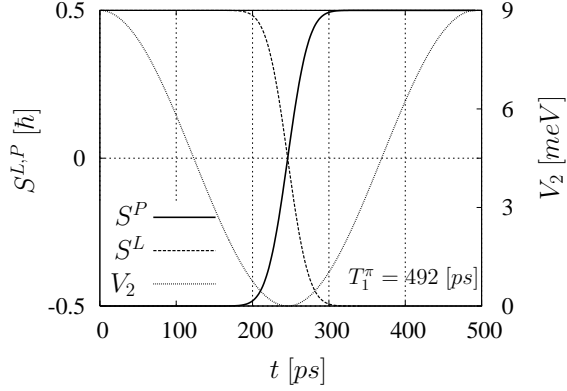}
\caption{Spin indices $S^P\equiv S^R$ for the right QD (solid curve)
and $S^L$ for the left QD (dashed curve),
and potential barrier height $V_2$ (dotted curve) as functions of time $t$
for duration time $T^{\pi}_1=492$ ps of the first $\pi$ pulse.}
\label{fig10}
\end{center}
\end{figure}

If the QD size is of the order of several ten nanometers,
the electrons are spatially separated even in the absence
of potential barrier \cite{Bednarek2003art}.
For the QDs of this size the Coulomb repulsion is sufficiently
strong so that the electrons are localized the QD boundaries
and form Wigner molecules \cite{Bednarek2003art}.
Only the localization of electrons
close to each other can lead to the rotation of their spins.
In the model of nanostructure (Fig. 9),
the effective width of the potential barrier is
much smaller than that of the double potential well.
Therefore, we can force the electrons to be localized
in the same region of the nanostructure, if we allow the parameter $V_2$
in Eq.~(\ref{parGauss}) to take on negative values during the time
evolution.  As a result, the potential barrier converts into
the potential well (Fig. 11)
and the electrons become localized in the same QD
for some time during the $\pi$ pulse duration time.

\begin{figure}[!ht]
\begin{center}
\includegraphics[width=0.45\textwidth]{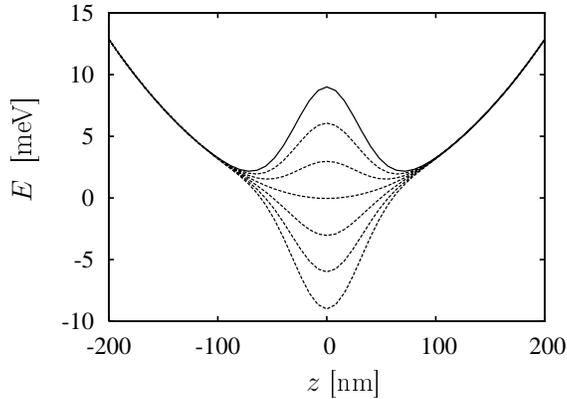}
\caption{Profile of confinement potential $E\equiv V$ [Eq.~(\ref{parGauss})]
as a function of coordinate $z\equiv x$ for $V_2 > 0$ and $V_2 <0$.
Shown are the plots for $V_2= 9,6,3,0,-3,-6,-9$ meV (from top
to bottom). At the initial time instant (solid curve),
$V_2=9$ meV, $d_B=60$ nm, $d_L=d_R=100$ nm.}
\label{fig11}
\end{center}
\end{figure}

In the first computer run, parameter $V_2$ was changed in a smooth manner
from 9 meV to $-3$ meV.  As expected, we obtained the considerably shorter
duration time of the spin rotation: the first $\pi$ pulse lasted merely for 46.1 ps,
i.e., it was more than ten times shorter than in the case of $V_2 \geq 0$.
It appears a question: to what extent can we shorten the duration time
of spin rotation by lowering the potential well bottom?
In order to answer this question, we have performed a series
of simulations, in which the minimum value $V_2^{min}$ of parameter $V_2$
was lowered from 0 to $-9$ meV.
In each simulation, the initial value of parameter $V_2$ was the same:
$V_2(t=0)=V_2^{max}=9$ meV.
The results are shown in Fig. 12.  According to our expectation,
the spin interchange duration time becomes shorter
if $V_2^{min}$ decreases.  However, we also observe an undesirable effect,
namely, the process ceases to be adiabatic, which results from the increasing
amplitude of changes of $V_2$ during time $T$.
The non-adiabaticity is visible in Figs. 12(a-d) as oscillations of energy difference
$\Delta E_{fi}(T)$ and deformations of plots $S_f^R(T)$ for small
duration time $T$.  The process can be treated
as adiabatic for $T \geq 25$ ps ($T \geq 55$ ps) if $V_2^{min}=0$ ($V_2^{min}=-9$ meV).
We conclude that the lowering of $V_2^{min}$ below $-3$ meV
is not advantageous.  The shortest time of the full interchange of spins
has been obtained for $V_2^{min}=-3$ meV.
Then, the orientation of spins was changed in the single swap
after time $T_1^{\pi}=46.1$ ps with accuracy 100 \%.

\begin{figure}[!ht]
\begin{center}
(a) \hspace{0.47\textwidth} (b) \\
\includegraphics[width=0.45\textwidth]{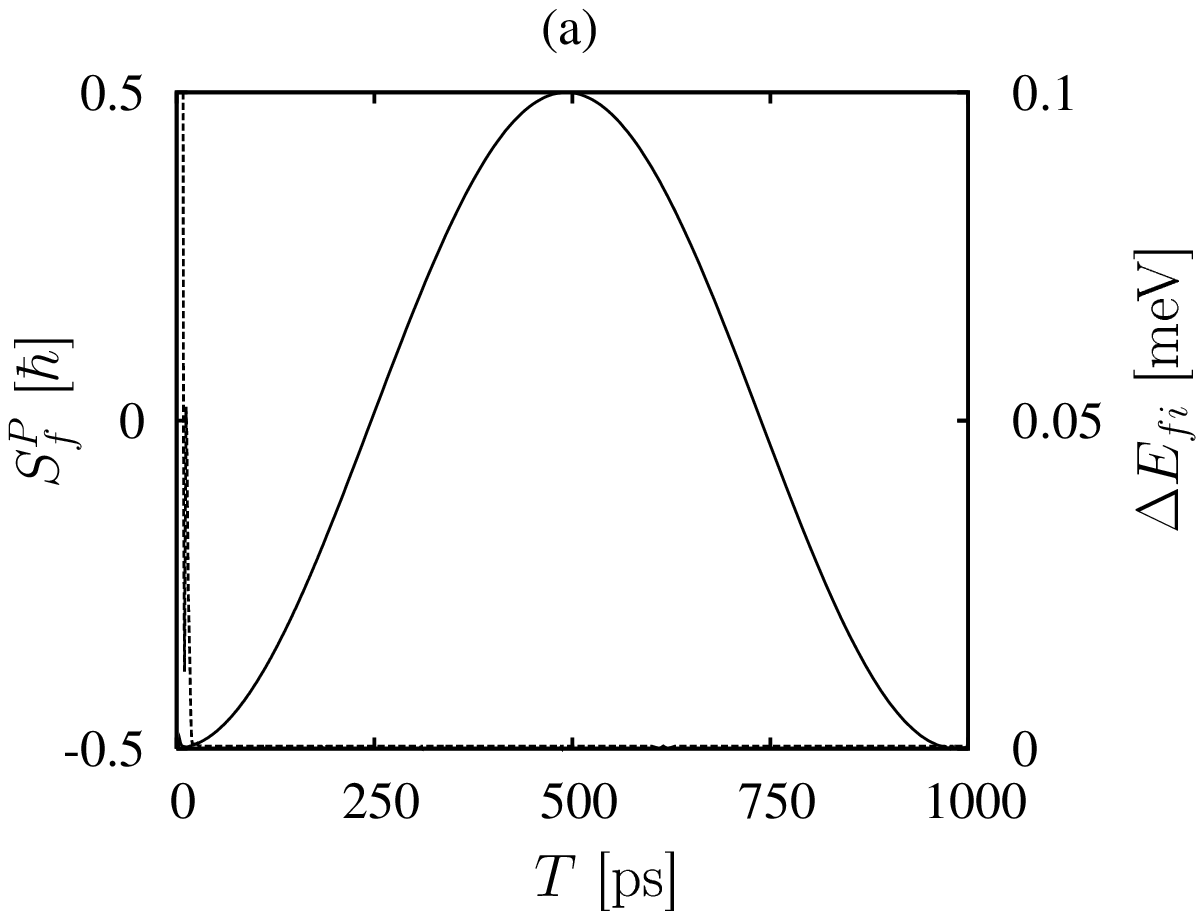}\hspace{0.08\textwidth}
\includegraphics[width=0.45\textwidth]{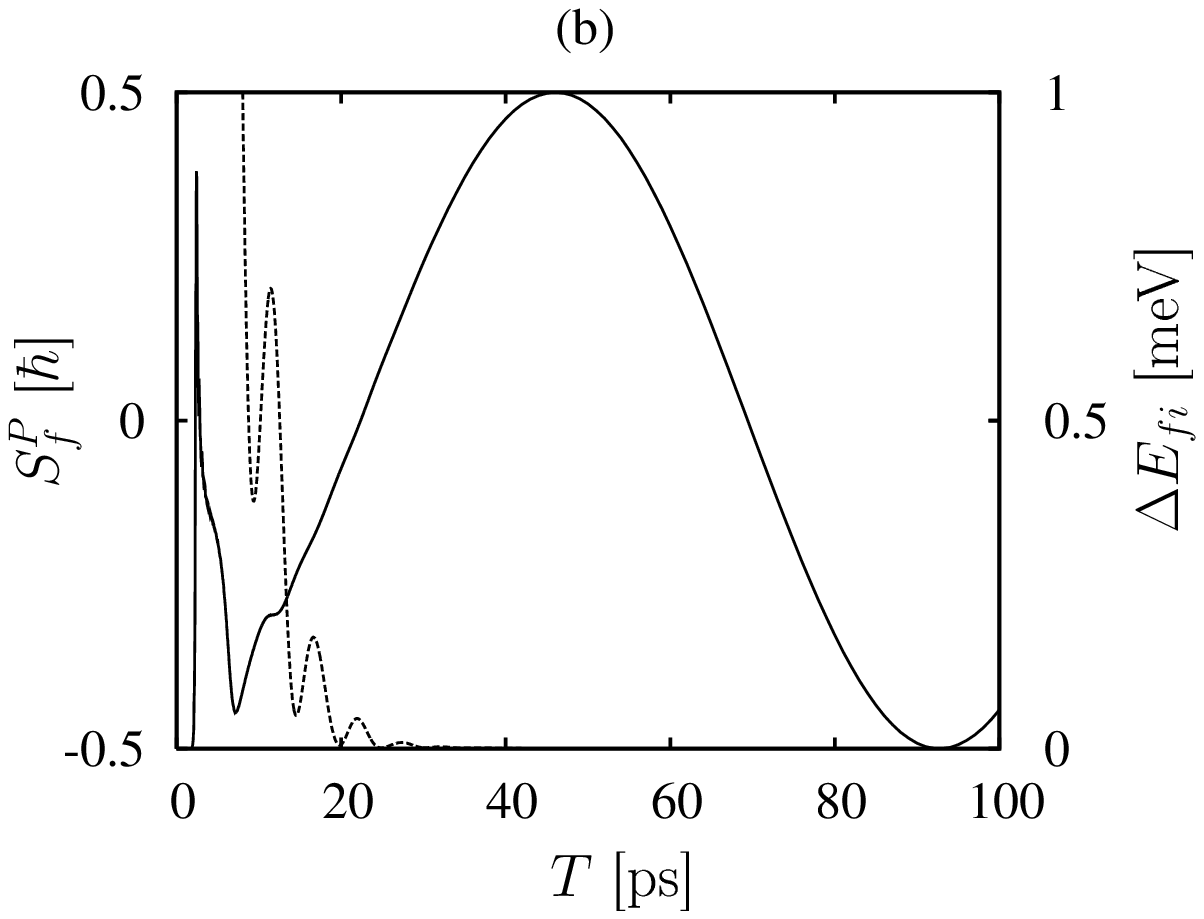}\\
\vspace{12pt}
(c) \hspace{0.47\textwidth} (d) \\
\includegraphics[width=0.45\textwidth]{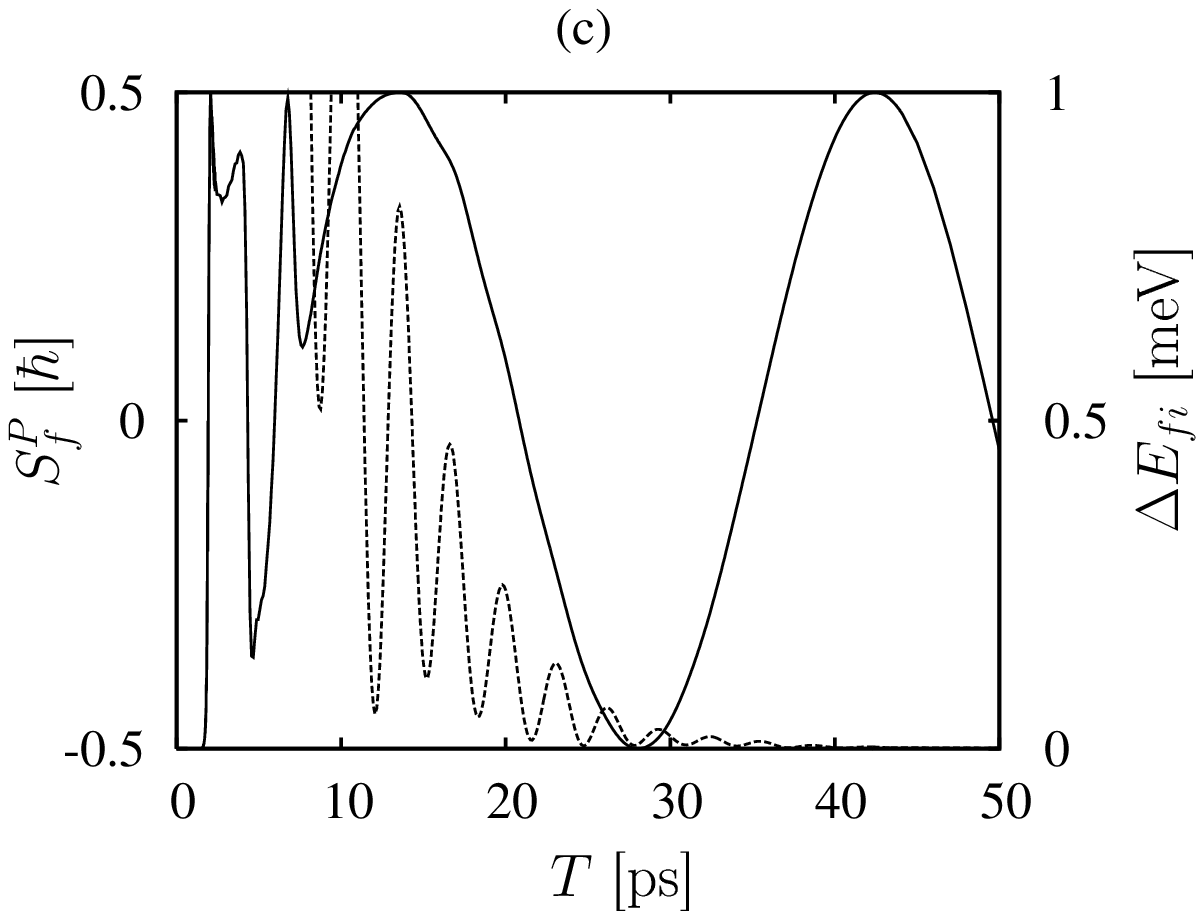}\hspace{0.08\textwidth}
\includegraphics[width=0.45\textwidth]{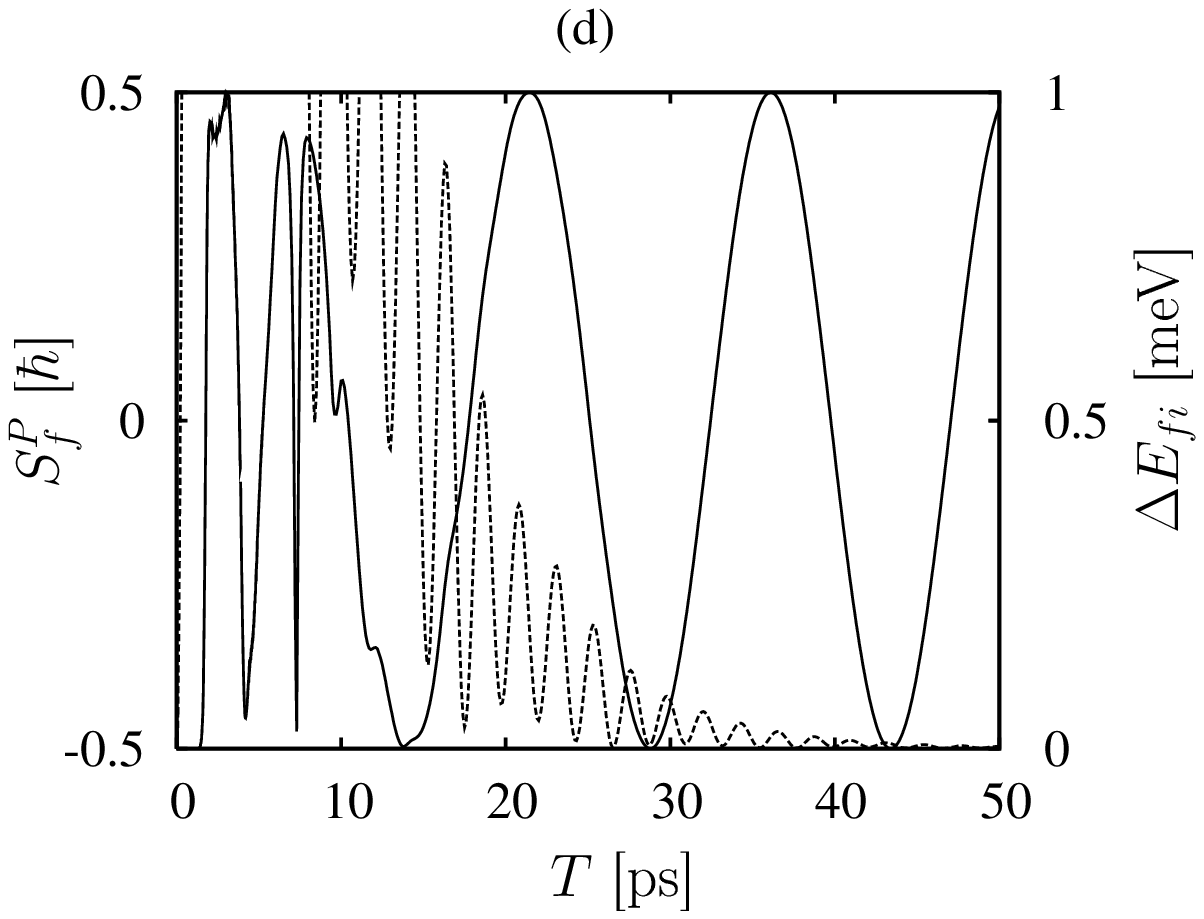}
\caption{Expectation value $S_f^P\equiv S_f^R$ of $z$ spin component in the right QD
(solid curve) and energy difference $\Delta E_{fi}$ (dashed curve)
between the final and initial states as a function of time duration $T$ of the process.
(a) $V_2^{min}=0$, (b) $V_2^{min}=-3$ meV, (c) $V_2^{min}=-6$ meV,
(d) $V_2^{min}=-9$ meV.}
\label{fig12}
\end{center}
\end{figure}

We note that for the laterally coupled QDs the change of
the interdot potential barrier into the deep potential well means
that a section of quantum wire is created below the gate electrodes,
which are responsible for the formation of electrostatic QDs \cite{Elzerman2003}.
However, in this process, the spin qubits can be destroyed if
the electrons will tunnel from the two-dimensional electron gas \cite{Elzerman2003}
to the QD.  Therefore, it seems that the mechanism of conversion of the
interdot potential barrier into the potential well will not lead
to the required rotation of spins in the laterally coupled QDs.
However, this mechanism can be applicable to the spin qubits in the vertically
coupled QDs \cite{Austing2004} and the coupled QDs formed in quantum wires
\cite{Bjork2002}.

\subsection{Asymmetric quantum dots}

The bias voltage applied to the coupled QDs leads to
the asymmetry of the confinement potential profile,
i.e., the potential well depths become different.
This suggests another mechanism of swapping the spins,
which can be modelled by raising and lowering
the bottom of one potential well.  Raising
the potential well bottom of one QD will stimulate
the flow of the electron to the second QD.
Therefore, both the electrons will be localized
in the same QD. The subsequent lowering of the potential
well bottom back to the initial level will cause
the flow of one of the electrons to the first QD.
If the duration time of changing
the potential well will be suitably adjusted,
the electrons will swap their spins.
This mechanism is advantageous since the electrons
meet in the same QD.  According to the previous results,
the stronger electron localization leads to
the shorter time of spin swapping.

In order to simulate this process,
we model the confinement potential by the linear combination
of two Gaussians, i.e.,
\begin{equation}
V(x)=V_1 \exp \left[ -(x-x_1)^2/d_1^2 \right]
+ V_2 \exp \left[ -(x-x_2)^2/d_2^2 \right] \;,
\label{Gauss2}
\end{equation}
where $V_1$ and $V_2$ are the potential well depths,
$x_1$ and $x_2$ are the positions of the centers
of the QDs, and $d_1$ and $d_2$ determine the sizes
of the QDs (Fig. 13).

\begin{figure}[!ht]
\begin{center}
\includegraphics[width=0.45\textwidth]{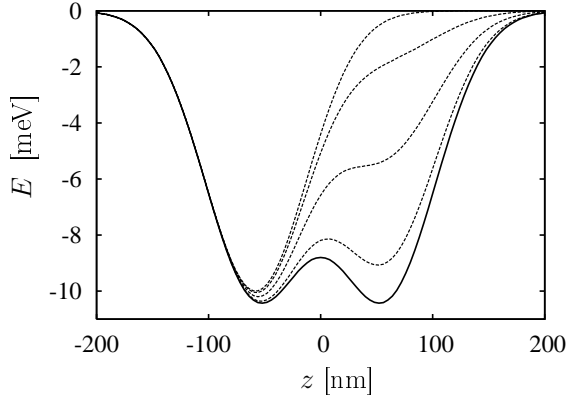}
\caption{Profile of confinement potential $E\equiv V$ [Eq.~(\ref{Gauss2})]
as a function of coordinate $z\equiv x$ for several time instants.
Parameter $V_1=-10$ meV is fixed.
At the initial moment (solid curve), $V_2(t=0)=-10$ meV.
Shown are also the potential profiles
for $V_2=-7,-5,-3$, and 0 meV.}
\label{fig13}
\end{center}
\end{figure}

We change parameter $V_2$ according to the smooth function of time
[Eq.~(\ref{cos})].  At the initial time, the electron in the right (left) QD
has spin $+\hbar/2$ ($-\hbar/2$).  We have performed several
simulations for different $\pi$ pulse duration time $T^{\pi}$
with the fixed parameters of confinement potential [Eq.~(\ref{Gauss2})]
$V_1=-10$ meV, $d_1=d_2=64$ nm, and $x_2=-x_1=58$ nm.
For $t=0$ (solid curve in Fig. 13) the bottom of both the potential wells
was at the level $-10.4$ meV, effective widths $d_1=d_2=48$ nm, $d_B=51$ nm,
and the relative height of the barrier was 1.7 meV.
In each computer run, the initial (minimum) value
of parameter $V_2$ was $V_2(t=0)=V_2^{min}=-10$ meV.
The potential profiles shown in Fig. 13 suggest that
it is not necessary to remove the right potential well
in order to localize both the electrons in the left QD.
Again, we are looking for the answer to question:
for which range of changes of the right potential well bottom
the duration time of interchanging the spins is the shortest?
The results of the simulations are displayed in Fig. 14.
As in the previous figures, for sake of clarity,
we present only the spin expectation values $S_f^R$
for the right QD, since $S_f^L$ can be obtained
by the reflection with respect to zero ordinate axis.
If parameter $V_2$ is changed from $V_2^{min}=-10$ meV to
$V_2^{max}=0$, the changes of the amplitude of the right potential well bottom
are so large that even for $T \simeq 100$ ps the spin interchange process
is non-adiabatic.  In this case, we observe the spin exchange
with accuracy 99.8 \% for $T_1^{\pi}=90.4$ ps.  The adiabatic process of
full interchange of spins is observed for $T$ of the order of several
hundred picoseconds.  Therefore, the process, during which one of the
potential wells is entirely removed, is not advantageous.
Therefore, we reduce the amplitude of changes of parameter $V_2$
by decreasing its maximum value $V_2^{max}$.
The useful results are obtained when raising the potential well bottom
to $V_2^{max}=-3$ meV [Fig. 14(a)].
However, we have to take into account that $V_2^{max}$ can not
be too small since then the potential well bottom
is not raised high enough so that the probability of electron tunneling
from the right to left QD is too small, which leads to
the long spin swap duration time.  For example, for $V_2^{max}=-7$ meV
[Fig. 14(c)] $T_1^{\pi}=713$ ps.  The shortest $\pi$ pulse time ($T_1^{\pi}=139$ ps)
has been obtained for $V_2^{max}=-5.5$ meV [cf. Fig. 14(b)].

 [Fig. 14(b)].

\begin{figure}[!ht]
\begin{center}
(a)\\
\includegraphics[width=0.45\textwidth]{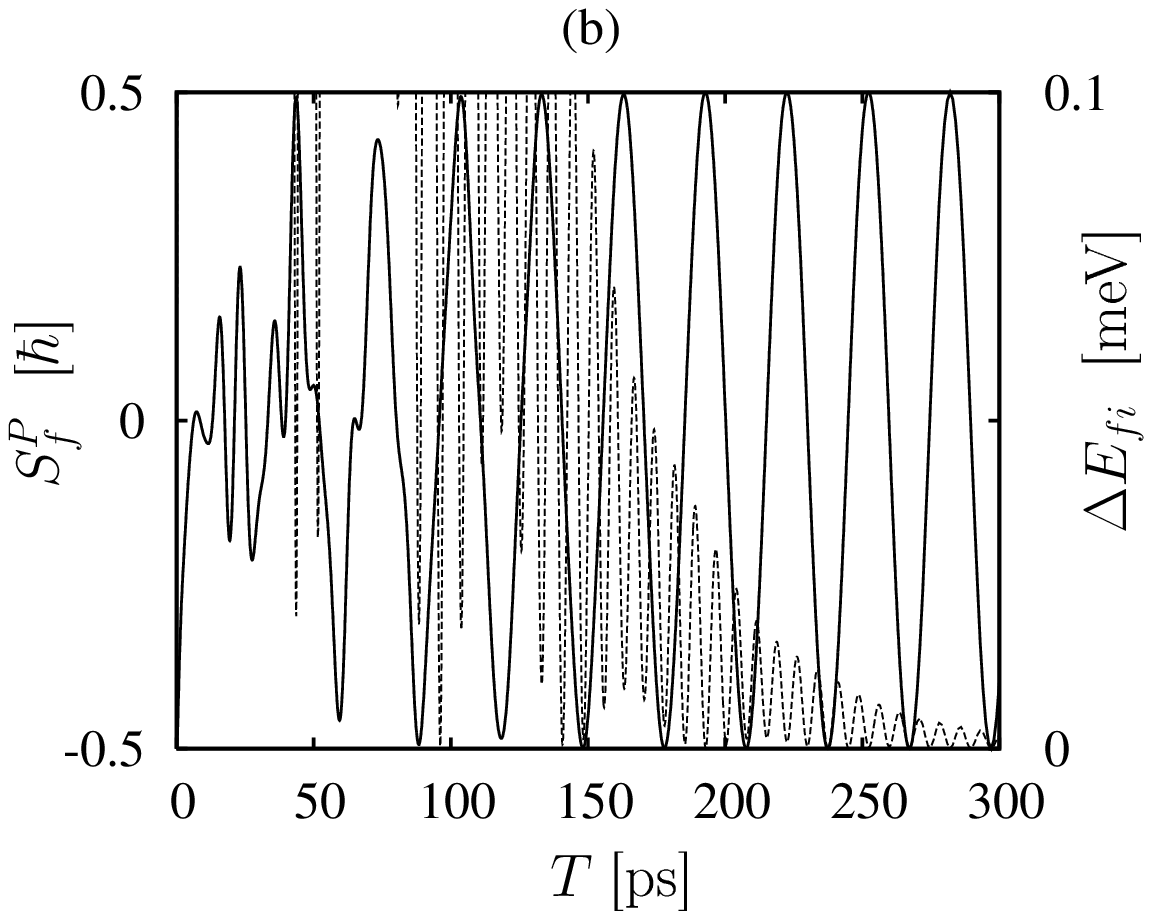}\\
\vspace{12pt}
(b) \hspace{0.47\textwidth} (c) \\
\includegraphics[width=0.45\textwidth]{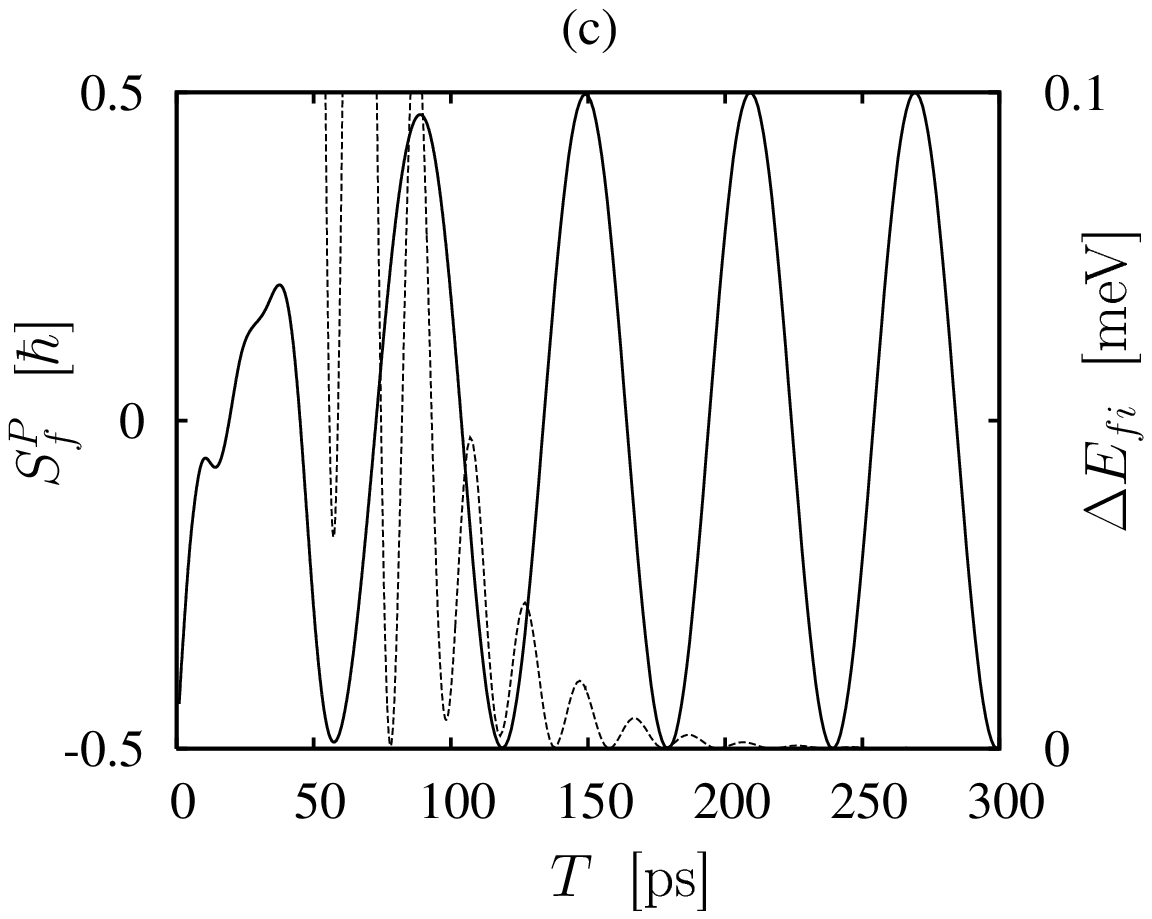}\hspace{0.08\textwidth}
\includegraphics[width=0.45\textwidth]{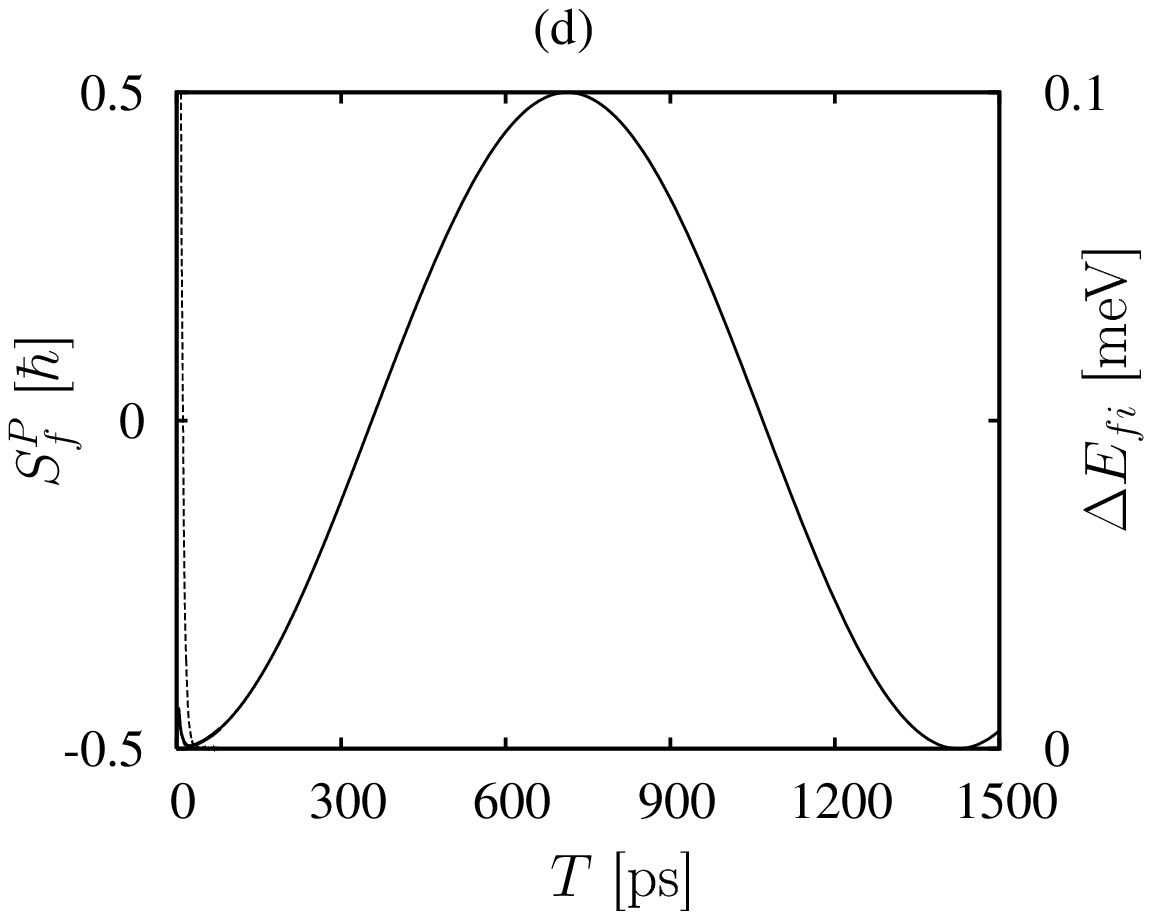}
\caption{Expectation value $S_f^P\equiv S_f^R$ (solid curve) of $z$ spin component
of the electron in the right QD and energy difference $\Delta E_{fi}$
(dotted curve) as functions of duration time $T$ of potential changes
for asymmetric QDs.
(a) $V_2^{max}=-3$ meV, (b) $V_2^{max}=-5$ meV, and (c) $V_2^{max}=-7$ meV.}
\label{fig14}
\end{center}
\end{figure}

We conclude that the time changes of the confinement potential,
which lead to asymmetry of QDs, result in the considerable shortening
of the spin swap time if the amplitude of potential well changes
is of the order of $\sim 5$ meV.

\section{Conclusions and Summary}

In the present paper, we have proposed a method for a theoretical
quantitative description of spin swapping process in a double QD.
Using this method we have estimated the spin swap duration time
induced by the exchange interaction between electrons
in vertically and laterally coupled QDs.  The results
obtained are also valid for quantum wires, which contain
coupled QDs.
The switching of the exchange interaction,
considered in the present paper, was triggered
by the corresponding time changes of the confinement potential profile.
The time dependence of the confinement
potential is of crucial importance in the process
of spin interchange.  We have shown that the confinement potential
should change in time in a smooth manner and sufficiently slowly
to ensure the adiabaticity of the spin swapping process.

We have also shown that the full spin interchange (with 100\% accuracy)
can hardly be achieved in the single spin swap process.
In the coupled QD system, the full spin interchange occurs as a result
of many-fold incomplete spin swaps.  The most promising
results, i.e., the shortest time of full spin swapping
in the adiabatic process, have been obtained when
smoothly changing in time the shape of the potential
in the interdot region from the potential barrier
to the potential well.  However, it seems to be
difficult to realize this process
experimentally in the same nanodevice by changing the voltages applied
to the gate electrodes, which define the laterally
coupled QDs.

The results obtained for asymmetric QDs
are rather disappointing: it turns out that
the asymmetry of the confinement potential
leads to the increase of the duration time of spin interchange,
since in this case the full spin swapping is obtained
after several incomplete spin swaps.
Both in the vertically and laterally coupled QDs
the spin swap duration time decreases
if  the range of the confinement potential becomes shorter.
This leads to the conclusion that the size of of coupled QD system
should be smaller than $\sim 100$ nm in order to obtain
the sufficiently short spin swap time.

In summary, the exchange-interaction induced processes
of spin rotation can serve as a mechanism to manipulate
with electron spin qubits in coupled QDs.  The spin swap duration time,
which is several orders of magnitude
shorter than the spin coherence time, can be obtained
for sufficiently small coupled QDs, in which the gate
voltages cause the smooth time changes of the confinement
potential.  A fabrication of the nanodevices,
which satisfy these requirements,
is a challenging task for the nanotechnology.

\end{document}